\documentclass[usenatbib,useAMS]{mnras}
\usepackage{graphicx}	
\usepackage{amsmath}	
\usepackage{amssymb}	
\usepackage{multicol}   
\usepackage{bm}		    
\usepackage{pdflscape}	
\usepackage{acro}
\usepackage[table,xcdraw, dvipsnames]{xcolor}

\newcommand{\D}{\ensuremath{\mathrm{\bm{D}}}}
\newcommand{\zcut}{\ensuremath{z_{\mathrm{cut}}}}
\newcommand{\HI}{\ensuremath{\mathrm{H}\scriptstyle\mathrm{I}}}
\newcommand{\Veff}{\ensuremath{V_{\mathrm{eff}}}}
\newcommand{\scatter}{\ensuremath{\sigma_{\mathrm{AM}}}}
\newcommand{\matched}{ALFALFA $\times$ NSA }

\newcommand{\pstrong}{\cellcolor[HTML]{B6DEB6}} 
\newcommand{\Pstrong}{\cellcolor[HTML]{4AAE4A}} 
\newcommand{\nstrong}{\cellcolor[HTML]{EEA8A8}} 
\newcommand{\Nstrong}{\cellcolor[HTML]{D62728}} 

\usepackage[utf8]{inputenc}
\usepackage[T1]{fontenc}
\usepackage{ae,aecompl}

\usepackage{newtxtext,newtxmath}

\DeclareAcronym{SHAM}{
  short = SHAM ,
  long  = subhalo abundance matching
}

\DeclareAcronym{AM}{
  short = AM ,
  long  = abundance matching
}

\DeclareAcronym{LF}{
  short = LF ,
  long  = stellar luminosity function
}

\DeclareAcronym{SMF}{
  short = SMF ,
  long  = stellar mass function
}

\DeclareAcronym{HIMF}{
  short = HIMF ,
  long  = $\HI$ mass function
}

\DeclareAcronym{BMF}{
  short = BMF ,
  long  = baryonic mass function
}

\DeclareAcronym{HMF}{
  short = HMF ,
  long  = halo mass function
}

\DeclareAcronym{NYU}{
  short = NYU-VAGC ,
  long  = New York University Value Added Galaxy Catalog
}

\DeclareAcronym{NSA}{
  short = NSA ,
  long  = Nasa Sloan Atlas
}

\DeclareAcronym{SDSS}{
    short = SDSS ,
    long  = Sloan Digital Sky Survey
}

\title[Dependence of AM on photometry and selection]
{The dependence of subhalo abundance matching on galaxy photometry and selection criteria}

\author[R.~Stiskalek, H.~Desmond, T.~Holvey and M.~G.~Jones]
{Richard Stiskalek$^{1, 2, 3}$\thanks{\href{mailto:richard.stiskalek@protonmail.com}{richard.stiskalek@protonmail.com}},
Harry Desmond$^1$\thanks{\href{mailto:harry.desmond@physics.ox.ac.uk}{harry.desmond@physics.ox.ac.uk}},
Thomas Holvey$^1$ and Michael G. Jones$^{4, 5}$
\\
$^{1}$Department of Physics, University of Oxford, Denys Wilkinson Building, Keble Road, Oxford OX1 3RH, UK\\
$^{2}$Fakultät für Physik, Ludwig-Maximilians-Universität München, 80333 München, Germany\\
$^{3}$School of Physics and Astronomy, University of Glasgow, Glasgow G12 8QQ, UK\\
$^{4}$Instituto de Astrofísica de Andalucía (IAA-CSIC), Glorieta de la Astronomía s/n, 18008, Granada, Spain\\
$^{5}$Steward Observatory, University of Arizona, 933 North Cherry Avenue, Rm. N204, Tucson, AZ 85721-0065, USA
}

\date{Last updated xxx Xxx x; in original form xxx Xxx x}

\pubyear{2021}

\begin{document}

\label{firstpage}
\pagerange{\pageref{firstpage}--\pageref{lastpage}}
\maketitle

\begin{abstract}
\Ac{SHAM} is a popular technique for assigning galaxy mass or luminosity to haloes produced in $N$-body simulations. The method works by matching the cumulative number functions of the galaxy and halo properties, and is therefore sensitive both to the precise definitions of those properties and to the selection criteria used to define the samples. Further dependence follows when \ac{SHAM} parameters are calibrated with galaxy clustering, which is known to depend strongly on the manner in which galaxies are selected. In this paper we introduce a new parametrisation for \ac{SHAM} and derive the best-fit \ac{SHAM} parameters as a function of various properties of the selection of the galaxy sample and of the photometric definition, including S\'{e}rsic vs Petrosian magnitudes, stellar masses vs $r$-band magnitudes and optical (SDSS) vs $\HI$ (ALFALFA) selection. In each case we calculate the models' goodness-of-fit to measurements of the projected two-point galaxy correlation function. In the optically-selected samples we find strong evidence that the scatter in the galaxy--halo connection increases towards the faint end, and that AM performs better with luminosity than stellar mass. The \ac{SHAM} parameters of optically- and $\HI$-selected galaxies are mutually exclusive, with the latter suggesting the importance of properties beyond halo mass. We provide best-fit parameters for the \ac{SHAM} galaxy--halo connection as a function of each of our input choices, extending the domain of validity of the model while reducing potential systematic error in its use.
\end{abstract}


\begingroup
\let\clearpage\relax
\endgroup
\newpage


\section{Introduction}

Currently, our best tool to understand the growth of structure in the Universe is simulation of dark matter halo formation and evolution. While cosmological hydrodynamical simulations have increased greatly in sophistication in the past decade, the complexity of baryonic processes means that simulations of the gravitational physics of dark matter only ($N$-body simulations) remain the most robust, with broad consensus on the resulting halo population across the range of algorithms that have been used~\citep{NFW,Knebe2011,Schneider}. To connect such simulations to observations, empirical and semi-analytic models have been developed to associate galaxies with the simulated haloes and track the coevolution of the two over cosmic time. These models vary considerably in complexity and scope, ranging from simple parametrised prescriptions for relating galaxy and halo properties (e.g. the Halo Occupation Distribution;~\citealt{Kravtsov_HOD, Zheng_HOD, Skibba}) to fully probabilistic models of galaxy formation as a function of halo mass accretion histories~\citep{EMERGE2018, Universe_machine, EMERGE}.

A model towards the empirical end of this spectrum is \ac{AM}, which is a simple and versatile prescription for the galaxy--halo connection. In its most basic form, \ac{AM} assumes a monotonic relation between halo mass and galaxy mass or luminosity, such that the brightest galaxy within a given survey volume is associated with the most massive halo produced in a similar volume of an $N$-body simulation, the next brightest with the next most massive and so on down the list~\citep{Kravtsov_HOD, Vale2004, Conroy2006, Behroozi2010, Moster2010}.
The method has subsequently been extended to incorporate subhaloes (\ac{SHAM}) and scatter in the relation between galaxies and haloes, and to vary the halo property used as the ``proxy'' for galaxy mass~\citep{Behroozi2010, Reddick2013, Chaves-Montero2016, Lehmann}. While reproducing the \ac{SMF} or \ac{LF} by construction, \ac{SHAM} has been shown to give good predictions for galaxy clustering and satellite fractions as a function of mass (e.g. ~\citealt{Conroy2006, Reddick2013, Lehmann}), and to be able to explain aspects of galaxy dynamics~\citep{Desmond_TFR, Desmond_FJR, Desmond_MDAR}. This suggests that \ac{SHAM} encapsulates important physics regarding the processes of galaxy formation, and also makes it useful for cosmological studies that require the galaxy--halo connection as input (e.g.~\citealt{Reddick_cosmo}).

With galaxy correlation function as the main observable that has been used so far to test and constrain \ac{SHAM}, the model is sensitive to effects that impact galaxy clustering. A systematic study of such effects on the results of \ac{SHAM} is however lacking. In particular, previous studies have taken for granted both survey selection criteria and photometric modelling choices, both of which are known to impact the inputs to \ac{SHAM}. The galaxy data most commonly used for \ac{SHAM} derives from the \ac{SDSS}\footnote{\url{https://www.sdss.org/}}, with previous studies typically preferring the Petrosian photometry pipeline from the \ac{NYU}\footnote{\url{http://sdss.physics.nyu.edu/vagc/}}~\citep{NYU}. The selection here is done mainly on the basis of $r$-band magnitude, tracing main sequence stars. Galaxies selected through emission at other wavelengths, e.g. $21\mathrm{cm}$ ($\HI$), are much more weakly clustered~\citep{Li,Martin,Papastergis_HIclust}. This requires \ac{SHAM} models designed for such samples to associate galaxies with a subset of the halo population significantly biased in quantities that affect clustering, for example formation time~\citep{Guo}.

The photometric pipeline used to reduce the raw images also has an impact on the luminosities and stellar masses derived. There has been much discussion in the literature as to the best method for determining these, both as regards the luminosity in the low surface brightness outer regions of galaxies~\citep{Blanton_subtraction, Bernardi, Kravtsov_SMF} and the method used to convert galaxy fluxes and spectra to stellar masses (\citealt{Baldry:2008ru} and references therein). Since these affect the \ac{LF} and \ac{SMF}, they alter the galaxy correlation functions produced by a given \ac{SHAM} model, and hence the best-fit parameters of \ac{SHAM} itself.

The stellar-to-halo mass relation has been extensively studied, with evidence accruing for an intrinsic scatter around $0.2~\mathrm{dex}$ (e.g.~\citet{Behroozi2010, Moster2010, Puebla2011}).
The stellar-to-halo mass relation of simulated faint galaxies was studied in~\citet{Munshi2021}, finding an increasing scatter in the relation towards the lower-mass halos.
Likewise, \citet{Redmapper} constrained the luminosity scatter at a fixed halo mass of red galaxies, finding similar values of scatter as in the stellar-to-halo-mass relation.
On the other hand, the relation between the gas mass of a galaxy and halo properties is relatively little constrained; there may even be a non-monotonic relation between the two (e.g.~\citealt{Guo}). This agrees with the direct measurement of the $M_{\HI}-M_h$ relation of~\citet{Guo2020}, who found that only the satellite $\HI$ mass increases monotonically with halo mass. \citet{Chauhan2020} also found a non-monotonic $M_{\HI}-M_h$ relation, and studied the physical drivers of the relation in a semi-analytic galaxy formation model. \citet{Lu2020} developed an empirical model for $\HI$ mass distribution within dark matter halos using galaxy group catalogues, and used it to predict $\HI$ masses and satellite fractions. A non-monotonic $M_{\HI}-M_h$ relation was also found in \citet{Calette}, who first impose a stellar mass--circular velocity relation via \ac{SHAM} and then use an empirical $\HI$--stellar mass relation to predict the galaxy $\HI$ content.

In this paper, we explore the impact of both photometric pipeline and selection criteria on the results and the best-fit parameters of \ac{SHAM}.
In particular, we will consider both optically selected galaxies from the \ac{NYU} and \ac{NSA}\footnote{\url{http://nsatlas.org/}}, and $\HI$-selected galaxies from the ALFALFA $\HI$ survey~\citep{Giovanelli_1, Giovanelli_2, Saintonge, Haynes_2018}\footnote{\url{http://egg.astro.cornell.edu/index.php/}}, in a matched catalogue. For the galaxy property used in \ac{SHAM}, we will consider both the S\'ersic and Petrosian (two-dimensional, i.e. elliptical, in the case of the \ac{NSA}) magnitudes along with their corresponding stellar mass estimates. S\'ersic magnitudes derive from fitting a S\'ersic profile~\citep{Sersic} to the surface brightness profiles of galaxies, while the Petrosian magnitude is the flux from within the Petrosian radius where the surface brightness is a fixed fraction of the mean surface brightness within that radius~\citep{Petrosian}. For each one we compute the luminosity and stellar mass functions, and the correlation functions in bins. We fit an \ac{SHAM} model in each case, deriving constraints on the \ac{SHAM} parameters and calculating the goodness-of-fit of the model by means of the Bayesian evidence. This will let us answer the following questions:

\begin{itemize}
    \item How do the best-fit \ac{SHAM} parameters vary with galaxy magnitude or stellar mass?
    \item Is there consistency between \ac{SHAM} results for different photometric pipelines and selection criteria; or, more generally, how do the inferred \ac{SHAM} parameters depend on these?
    \item Which of the above is most suitable for doing \ac{SHAM} in the sense of maximising goodness-of-fit of the model?
    \item What are the differences between \ac{SHAM} based on $M_r$ vs $M_*$, and for which does \ac{SHAM} provide the best fit?
    \item How do the two-point correlation functions, \acp{LF} and, \acp{SMF} differ between $r$-band and HI-selected samples, and how does this impact the results of \ac{SHAM}?
\end{itemize}

Our aim is therefore to determine both how well \ac{SHAM} works under various conditions, and the sensitivity of its parameters to galaxy sample and systematic effects in galaxy modelling. We anticipate this to prove useful for understanding the physical significance of the model, and for ensuring that it is as accurate as possible for specific science applications. We work in the context of a custom-built \ac{SHAM} model that is however similar to those in the literature (especially~\citealt{Lehmann}), as we describe fully in Sec.~\ref{sec:catalogs}.
Within the \ac{SHAM} framework, we match galaxies to both haloes and subhaloes. Unless an explicit distinction is drawn, we refer to both haloes and subhaloes collectively as ``haloes''. We assume a flat $\Lambda$CDM model with a matter density parameter of $\Omega_m = 0.295$ and quote results assuming a Hubble constant $H_0 = 70~\mathrm{km}~ \mathrm{s}^{-1}~\mathrm{Mpc}^{-1}$, unless $h \equiv H_0 / (100~\mathrm{km}~ \mathrm{s}^{-1}~\mathrm{Mpc}^{-1})$ is explicitly mentioned. All logarithms are base-10. We make our code publicly available.\footnote{\href{https://github.com/Richard-Sti/ClusterSHAM}{github.com/Richard-Sti/ClusterSHAM}}

The structure of the paper is as follows. Sec.~\ref{sec:data} documents the observed and simulated data that we use. Sec.~\ref{sec:method} describes our methodology, including \ac{SHAM} modelling, calculation of correlation functions, and the likelihood framework for inferring \ac{SHAM} parameters and testing goodness-of-fit. Sec.~\ref{sec:results} presents our results, Sec.~\ref{sec:discussion} discusses the broader impact of our study and suggests avenues for further work, and finally Sec.~\ref{sec:conc} concludes.

\section{Observed and simulated data}
\label{sec:data}

\subsection{Observational data}\label{sec:data_obs}

We study three galaxy catalogues. The first, the \ac{NYU}\footnote{\url{http://sdss.physics.nyu.edu/vagc/}}, is based on the \ac{SDSS} DR7~\citep{SDSS, SDSS_DR7, SDSStargeting}, consisting of $7966~\text{deg}^2$ of spectroscopic coverage. We use the \ac{NYU} Petrosian model magnitudes K-corrected to $z=0.0$~\citep{Kcorrect}, and select galaxies with apparent \textit{r}-band magnitude $10 < m_r < 17.7$ and redshift $0.01 < z < 0.15$ from the \ac{NYU} large-scale structure sample.\footnote{\url{http://sdss.physics.nyu.edu/vagc/lss.html}} The lower redshift limit excludes nearby galaxies whose redshift estimate may be dominated by peculiar velocities. Despite the fact that \ac{NYU} extends significantly beyond $z=0.15$ we opt for this upper limit to ensure fair comparison with the \ac{NSA}, which only extends to $z=0.15$.

The second catalogue we use is the \ac{NSA}, a database of images and parameters of local galaxies based primarily on results from the \ac{SDSS} DR13~\citep{SDSS_DR13} and the Galaxy Evolution Explorer~\citep{GALEX}. v1\_0\_1 of the NSA\footnote{\url{https://www.sdss.org/dr13/manga/manga-target-selection/nsa/}} contains $641,409$ galaxies up to $z=0.15$. The \ac{NSA} includes elliptical Petrosian and S\'ersic aperture photometry fits K-corrected to $z=0.0$, both of which we use in our analysis. Based on simulations, the elliptical Petrosian photometry avoids biases present in the S\'ersic photometry and is, therefore, considered more reliable. We again limit the apparent \textit{r}-band magnitude to $10 < m_r < 17.6$ and use the same redshift cuts as for the \ac{NYU}. Furthermore, we only select galaxies which are also present in the \ac{NYU} large-scale structure sample.
Both the \ac{NYU} and \ac{NSA} galaxy catalogues are corrected for fibre collisions: fibre collided galaxies, which would otherwise lack redshifts, are assigned the nearest neighbour's redshift~\citep{SDSS_early, Zehavi_Fiber}.

Our final galaxy catalogue consists of optically-selected \ac{NSA} \emph{and} $\HI$-selected galaxies from ALFALFA, a blind, second generation extragalactic $\HI$ survey. The full ALFALFA catalogue, denoted $\alpha$.100, contains $\sim$31,500 $\HI$ line sources up to $z=0.06$ with $\HI$ masses ranging from $10^6$ to $10^{10.8}~M_\odot$~\citep{Haynes_2018}. We include both code 1 and code 2 ALFALFA sources; although the completeness of the survey is different between these, our method for determining the correlation functions (Sec.~\ref{sec:CF}) applies equally to both of them. ALFALFA has a partial overlap with \ac{SDSS}, allowing for matching between the two surveys. We begin with the SDSS match performed in~\citet{ALFALFA-SDSS}. We then match the position of the optical counterpart to the sources in the Nasa Sloan Atlas using an on-sky angle tolerance of $5\arcsec$ and a line-of-sight distance tolerance of 10 Mpc. These fairly stringent criteria yield a low probability of mismatches and hence a high sample purity, while still retaining $21,776$ galaxies. We restrict the footprint of the combined catalogue to the ``Spring'' section of the ALFALFA footprint\footnote{\url{http://egg.astro.cornell.edu/alfalfa/scheds/status\_spr11.php}} to avoid regions of poor coverage in SDSS, which removes $\sim$15\% of the total survey area. Of the remaining $\HI$ sources with optical counterparts, $83.6\%$ have stellar mass estimates in the NSA ($9,802$ galaxies). The missing galaxies are mainly nearby and with low $\HI$ mass. We will locate galaxies in the matched catalogue using the sky coordinates and redshift quoted in the NSA.

\subsection{Simulation data}\label{sec:data_sim}

We base our \ac{SHAM} mock catalogues on the DarkSky simulation\footnote{\url{https://darksky.slac.stanford.edu/}}~\citep{DarkSky}. In particular we use the DarkSky $400~\text{Mpc}/h$ box (ds14\_i\_4096) run with \textsc{2hot} code~\citep{2HOT} and $4,096^3$ particles. The particle mass is $7.63\times 10^7~M_\odot/h$, and the minimum halo virial mass in the catalogue is $3 \times 10^9~M_\odot/h$, corresponding to $\sim 40$ particles. In the subsequent analysis we ensure that the haloes matched to galaxies typically contain at least $200$ particles to minimise the potential bias induced by matching poorly-resolved haloes to the faintest galaxies (see also Sec.~\ref{sec:systematic_uncertanties}). Haloes and subhaloes are identified using \textsc{rockstar} halo finder~\citep{Behroozi_Rockstar} and the \textsc{consistent trees} merger tree builder~\citep{Behroozi_Consistent}. Halo virial masses are calculated as regions with overdensity $\Delta_{\text{vir}}=178$, following~\cite{Bryan_Norman}. DarkSky assumes a flat $\Lambda$CDM cosmology with $H_0 = 68.8~\mathrm{km}~ \mathrm{s}^{-1}~\mathrm{Mpc}^{-1}$, $\Omega_m = 0.295$, scalar spectral index $n_s = 0.968$, and root-mean-square matter fluctuation on 8 Mpc/$h$ scales $\sigma_8 = 0.834$.

\section{Methodology}\label{sec:method}

\subsection{Subhalo abundance matching}\label{sec:catalogs}

The fundamental assumption of \ac{SHAM} is that there exists a near-monotonic relation between a galaxy property $\mathfrak{g}$ (typically luminosity or stellar mass) and a halo property $\mathfrak{h}$ (typically a function of virial mass and concentration):
\begin{equation}\label{eq:abundance_matching}
    N_{\mathrm{gal}} \left(\bar{\mathfrak{g}}\right)
    =
    N_{\mathrm{halo}} \left(\bar{\mathfrak{h}}\right),
\end{equation}
where $N_{\mathrm{gal}} \left(\bar{\mathfrak{g}}\right)$ is the average number density of galaxies with $\mathfrak{g}>\bar{\mathfrak{g}}$ and $N_{\mathrm{halo}} \left(\bar{\mathfrak{h}}\right)$ is the average number density of haloes with $\mathfrak{h}>\bar{\mathfrak{h}}$. $\mathfrak{h}$ is referred to as the \emph{halo proxy} for $\mathfrak{g}$.

We present a new \ac{SHAM} model that uses a continuous parameter to describe the halo proxy, interpolating between the peak halo virial mass and present-day virial mass. This is a modification of the proxy introduced in~\cite{Lehmann}, which interpolates between the halo virial mass and maximum circular velocity at the time of the peak halo mass.

\begin{figure}
    \centering
    \includegraphics[width=1.0\columnwidth]{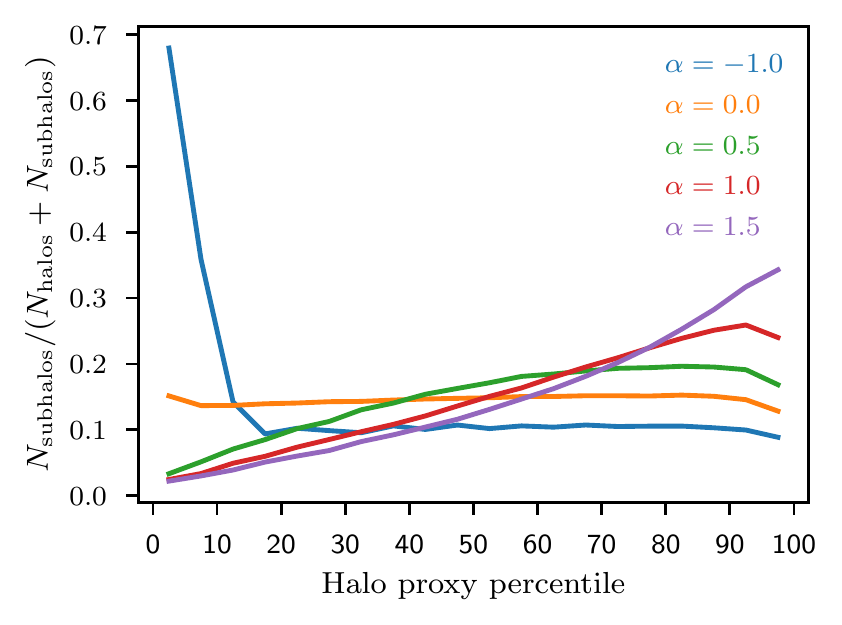}
    \caption{The proportion of subhaloes in the halo proxy list defined by Eq.~\eqref{eq:halo_proxy} for different choices of $\alpha$, percentile binned by the halo proxy ranking. When matching the haloes and subhaloes to galaxies, the most massive galaxy is assigned to the top-ranked halo, and so on down the list. Choosing $\alpha > 0$ boosts the proportion of upper-ranked subhaloes which are then matched to galaxies.}
    \label{fig:subhalo_fractions}
\end{figure}

The model has two free parameters, a generalised halo proxy $m_{\alpha}$ and a Gaussian scatter $\scatter$. The halo proxy is defined to be an interpolation between the peak virial mass over the history of the halo, $M_\text{peak}$, and the present-day value $M_0$:
\begin{equation}\label{eq:halo_proxy}
    m_{\alpha}
    =
    M_0
    \left( \frac{M_{\mathrm{peak}}}{M_0} \right)^{\alpha}.
\end{equation}
$m_{\alpha}$ then provides a smooth, continuously varying dependence on the two halo properties as a function of $\alpha$. If $\alpha = 0$ haloes are ranked simply by their present virial mass, and if $\alpha = 1$ by their peak virial mass. Ranking haloes by their peak virial mass ($\alpha = 1$) produces a stronger clustering signal than the present virial mass and increasing $\alpha$ further would further boost the clustering. It is a-priori expected that ranking haloes by $M_\text{peak}$ ought to be preferable to using $M_0$~\citep{Conroy2006}. Haloes with peak virial mass equal to their present-time virial mass are insensitive to $\alpha$, so that tuning $\alpha$ directly controls the rank of haloes that have undergone mass stripping (typically subhaloes). For $\alpha > 0$ haloes that have peaked in mass in the past are ranked higher in the halo proxy list, whereas for $\alpha < 0$ such haloes are ranked lower. Therefore, $\alpha$ can be interpreted as controlling the fraction of simulated subhaloes matched to galaxies.
We demonstrate this in Fig.~\ref{fig:subhalo_fractions}, which shows the dependence of the subhalo fraction on $\alpha$ in percentile bins.

The scatter in \ac{SHAM}, parametrised by $\scatter$, arises because the baryon content of a halo depends on parameters other than its mass, with the majority of scatter deriving from the halo mass accretion history~\citep{Tinker:2016zpi}. We implement the scatter using the deconvolution method described in~\citet{Behroozi2010}, and treat it as a free parameter. For $\scatter = 0$, abundance matching produces a one-to-one relation between the galaxy and halo proxy satisfying Eq.~\eqref{eq:abundance_matching}. If $\scatter \neq 0$, then the galaxies matched initially at $\scatter = 0$ are up- or down-scattered on the discrete galaxy-halo ladder. Non-zero values of $\scatter$ therefore add scatter to the galaxy-halo connection without affecting either the galaxy or halo proxy number densities themselves. It should be noted that our \ac{SHAM} model does not explicitly account for the uncertainties in the measured $\mathfrak{g}$ values (luminosity of mass), so that $\scatter$ contains both intrinsic and observational contributions.

We will also consider the possibility that the galaxies in our sample form in some subset of the total halo population. This is particularly relevant to HI-selected samples, which have overall significantly lower clustering than optically-selected samples~\citep{Guo, Papastergis_HIclust}. In particular, we consider pre-selection according to halo formation time, retaining only the haloes in the simulation with a peak-mass redshift lower than some cut-off $z_\mathrm{cut}$, while ranking the remaining haloes by their present mass.


\subsection{The galaxy luminosity and mass functions}\label{sec:LF}

\Acl{SHAM} requires as input the abundance functions of $\mathfrak{g}$ in the observational data and $\mathfrak{h}$ in the simulation. To calculate the luminosity and stellar mass functions for the optically selected samples we use the $1 / V_{\max}$ method~\citep{Schidt:1968}. This is a simple, non-parametric technique correcting for Malmquist bias, whereby intrinsically brighter galaxies are detected to larger distances. In this method, each galaxy is weighted by $V_{\max}$, the maximum comoving volume in which it could be located and still be detected by the survey, given its intrinsic brightness and the survey's limiting apparent magnitude threshold~\citep{Malmquist1920, Malmquist1922}. Following the standard definition of apparent magnitude, the maximum redshift $z_{\max}$ at which $i$-th galaxy of absolute magnitude $M_i$ can be located is calculated by solving the equation:
\begin{equation}\label{eq:apparent_magnitude_zmax}
    m_{\mathrm{lim}}
    =
    M_i + 25 + 5\log D_{\mathrm{L}}(z_{\max}) + K_i(z_{\max}),
\end{equation}
where $m_{\mathrm{lim}}$ is the limiting apparent magnitude of the survey and $D_{\mathrm{L}}(z_{\max})$ is the luminosity distance corresponding to $z_{\max}$~\citep{Hogg1999}. $K_i (z_{\max})$ is the estimated K-correction coefficient of the $i$-th galaxy if it were located at $z_{\max}$. We fit a simple multi-layer perceptron algorithm (see e.g.~\citet{10.5555/525960}) to predict the K-correction coefficient from the galaxy redshift, mass-to-light ratio, metallicity, and stellar formation rate such that $K_i (z_{\max})$ is the network's output when the observed galaxy redshift is replaced with $z_{\max}$. A multi-layer perceptron consists of input and output layers of neurons, with possibly several intermediate hidden layers. The input layer contains one neuron per input feature and the output layer consists of a single neuron, producing a scalar output. The number of neurons and hidden layers is adjustable. Each neuron in the hidden layers and output layer is passed a transformed weighted sum of all outputs from the previous layer. By adjusting the weights the network learns the mapping from a set of features to a scalar. In this particular problem, the network shows little sensitivity to the number of hidden layers and we choose 3 hidden layers containing 8, 4, and 2 neurons. We use the \textsc{scikit-learn}\footnote{\url{https://scikit-learn.org/stable/modules/neural_networks_supervised.html}} implementation of the perceptron network.

Eq.~\eqref{eq:apparent_magnitude_zmax} can also be solved for $z_{\min}$, the minimum redshift at which a galaxy would have been observed by the survey given a bright-end limit on apparent magnitude. As our optically-selected catalogues are ultimately volume-limited, values of $z_{\max}$ above the survey limiting redshift are set to to the limiting redshift (Sec.~\ref{sec:data_obs}).

With $D_{\mathrm{c}}(z)$ the comoving distance as a function of cosmological redshift, the maximum possible comoving volume for each galaxy is simply calculated under the assumption of a flat Universe ($\Omega_k = 0$):
\begin{equation}\label{eq:vmax}
    V_{\max}
    =
    \frac{\Omega_{\mathrm{S}}} {3}
    \left[ D_\mathrm{c} (z_{\max})^3 - D_\mathrm{c} (z_{\min})^3 \right],
\end{equation}
where $\Omega_{\mathrm{S}}$ is the solid angle spanned by the survey. $\Omega_{\mathrm{S}} = 7966~\deg^2$ for the \ac{NYU} and \ac{NSA} large-scale structure samples. The luminosity function $\Phi(M_r)$ is then obtained as a sum over the $1/V_{\max}$ contributions from the $N$ galaxies falling within the $j$-th absolute magnitude bin of width $\mathrm{d} M_r$:
\begin{equation}\label{eq:differential_luminosity_function}
    \Phi_j \: \mathrm{d} M_r
    =
    \sum_{i=1}^{N} \frac{1}{V_{{\max}_i}}.
\end{equation}
The \acl{SMF} $\Phi (\log M_*/M_\odot)$ is calculated analogously, using the same $1/V_{\max}$ coefficients but instead binning the galaxies by $\log M_*/M_\odot$. To avoid biased estimates, each survey must be complete to the chosen magnitude limit and contain no notably under- or overdense regions. To test this we apply the $V/V_{\max}$ test by calculating the $V/V_{\max}$ ratio for each galaxy, where $V$ is the volume given by the redshift at which the galaxy was actually detected~\citep{Schidt:1968}. If the survey is complete and the sources are uniformly distributed, then $V / V_{\max}$ should be uniformly distributed from $0$ to $1$, which is true for both the \ac{NYU} and \ac{NSA} samples.

\begin{figure*}
    \centering
    \includegraphics[width=1.0\textwidth]{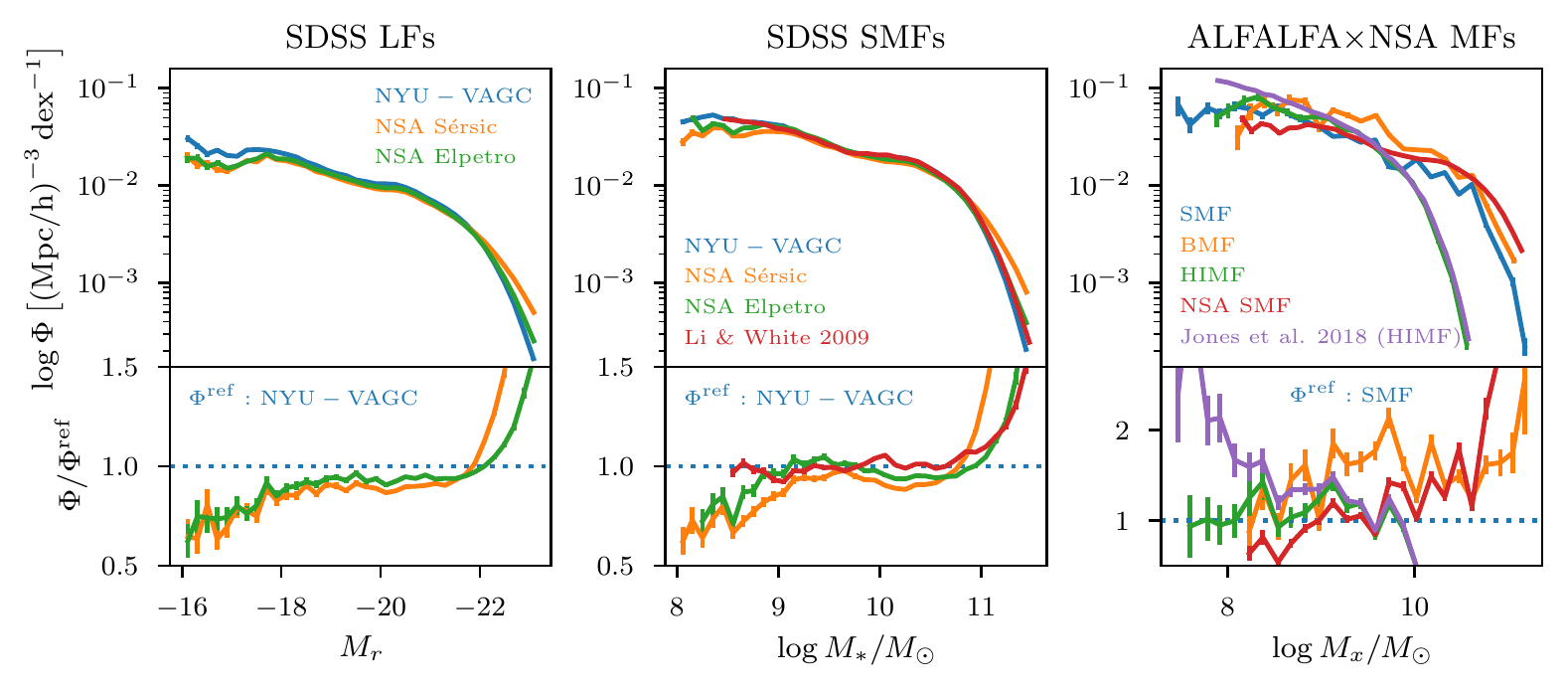}
    \caption{
    \emph{Left:} $r$-band \aclp{LF} for the \ac{NYU} Petrosian, \ac{NSA} S\'ersic and \ac{NSA} elliptical Petrosian catalogues. \emph{Centre:} \aclp{SMF} for the \ac{NYU} Petrosian, \ac{NSA} S\'ersic and \ac{NSA} elliptical Petrosian catalogues along with the \ac{SMF} from~\citet{Li2009} for comparison, which comes from a $z < 0.5$ \ac{NYU} sample. \emph{Right:} \acl{SMF}, \acl{BMF} and \acl{HIMF} for the \matched catalogue and comparison to \ac{NSA} elliptical Petrosian \ac{SMF} and the \ac{HIMF} of~\citet{Jones2018}, which is based on the full ALFALFA Spring sample. The subscript $x$ indexes the mass definition and the error bars represent the standard Poisson counting error.}
    \label{fig:LFs_MFs}
\end{figure*}

The resulting \acp{LF} and \acp{SMF} for the optically-selected catalogues are shown in the first two panels of Fig.~\ref{fig:LFs_MFs}, where the residuals are taken with respect to the \ac{NYU} \ac{LF} or \ac{SMF}. In particular, both \ac{NYU} Petrosian and \ac{NSA} elliptical Petrosian \acp{LF} and \acp{SMF} display similar number densities as expected. The \ac{NSA} S\'ersic \ac{LF} and \ac{SMF} are significantly different from the \ac{NYU} counterparts in predicting more galaxies at the bright end and fewer at the faint end. This is because the NSA photometric reduction includes more light in galaxies' low surface brightness wings \citep{Blanton_subtraction}.

We use ALFALFA $\HI$ masses ($M_{\HI}$) and \ac{NSA} elliptical Petrosian stellar masses to calculate the stellar, baryonic, and $\HI$ mass functions for our matched catalogue. The baryonic masses are estimated as
\begin{equation}\label{eq:baryonic_mass}
    M_{\mathrm{B}} = M_* + 1.4 M_{\HI},
\end{equation}
where the factor of $1.4$ accounts for the presence of cosmic helium, as Big Bang nucleosynthesis produces $\sim 25\%$ of He by mass~\citep{Nucleosynthesis}. We calculate the effective volume $\Veff$ for each galaxy in which it could have been located and still be observed by the survey following~\citet{Zwaan}. The $\Veff$s are the maximum-likelihood analogues of $V_{\max}$ values for optical-surveys, however calculated using the ALFALFA $\HI$ mass and velocity width completeness. As described in Sec.~\ref{sec:data}, we find that the matching process systematically eliminates nearby, low-$M_{\HI}$ galaxies. To account for this, we bin the galaxies in distance and $\HI$ mass, and calculate the proportion of galaxies in each bin with stellar mass estimates. We then multiply each galaxy's $\Veff$ value by fraction corresponding to the bin the galaxy is in. The resulting $\Veff$ estimates are then used to calculate the differential mass functions according to Eq.~\eqref{eq:differential_luminosity_function}.

\begin{figure}
    \centering
    \includegraphics[width=1.0\columnwidth]{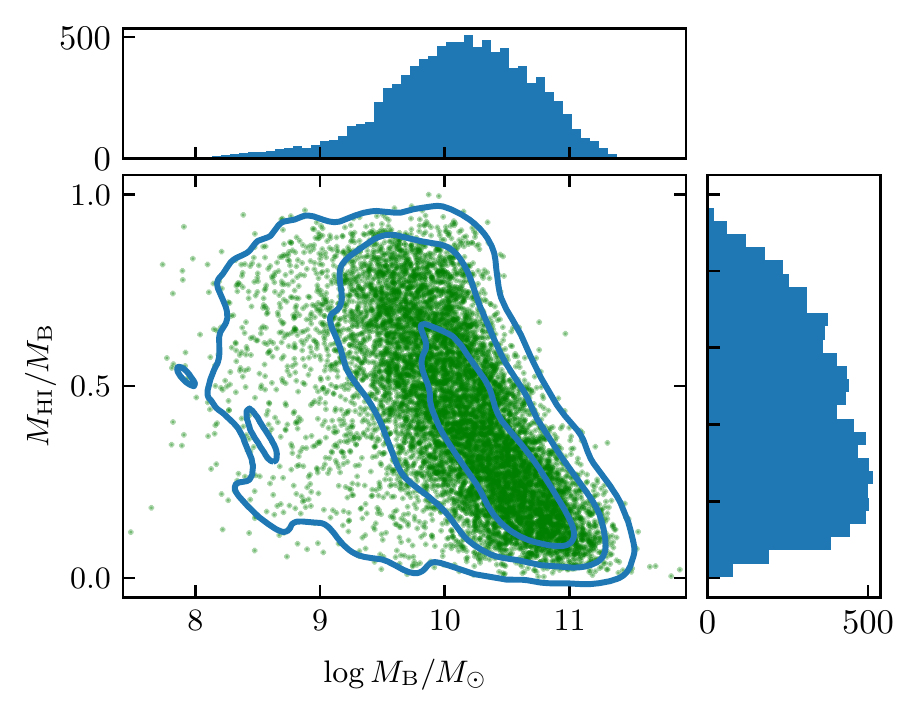}
    \caption{Gas fractions of galaxies in the \matched catalogue as a function of baryonic mass. The contours show the minimal areas enclosing $39\%$, $86\%$ and $99\%$ of the sample.}
    \label{fig:gas_fraction}
\end{figure}

In the third panel of Fig.~\ref{fig:LFs_MFs} we show the resulting \acl{SMF} (SMF), \acl{BMF} (BMF) and \acl{HIMF} (HIMF) for galaxies in the matched catalogue calculated using the $\Veff$ values. The residuals are taken with respect to the matched catalogue's \ac{SMF} and for comparison we also show the \ac{NSA} elliptical Petrosian \ac{SMF} (optical selection). The $\HI$-selected \ac{SMF} drops off significantly faster than its optically-selected counterpart. On the other hand, the \ac{BMF} is boosted relative to both the $\HI$-selected \ac{SMF} and \ac{HIMF} and partially overlaps with the \ac{NSA} elliptical Petrosian \ac{SMF}. Compared to~\citet{Jones2018} our \ac{HIMF} (and consequently \ac{BMF}) begins to turn over at the faint end: we discuss this in Sec.~\ref{sec:systematic_uncertanties}. In the matched catalogue there is a significant number of high baryonic mass galaxies that have non-negligible gas content, therefore explaining why the \ac{BMF} number density is higher than \ac{SMF} at the massive end of the mass function. The gas fractions are shown in Fig.~\ref{fig:gas_fraction}.

\subsection{Clustering measurements}\label{sec:CF}

To assess the extent to which the clustering of the \ac{SHAM} mock catalogues resembles that of the real Universe we compute the projected two-point correlation function, $w_p$. This is defined as the integral of the full 3D galaxy correlation function $\xi$ along the line of sight. Denoting separation along the line of sight $\upi$ and perpendicular to the line of sight $r_p$, the projected two-point correlation function is calculated as
\begin{equation}
    w_p (r_p)
    =
    2 \int_{0}^{\upi_{\max}} \xi (r_p, \upi ) \mathrm{d} \upi.
\end{equation}
In theory, the upper integration limit should be $\upi_{\max} \to \infty$ for $w_p(r_p)$ to be a purely real-space quantity. However, due to the finite size of the simulation or survey volume arbitrarily separated pairs cannot be counted. Moreover, in practise distant pairs will show little correlation over the scales which we are interested in. Therefore, it is sufficient to pick some finite integration limit $\upi_{\max}$, for which we choose $60~\mathrm{Mpc}/h$. This introduces a residual dependence on redshift-space distortions (RSD) when $\upi_{\max}$ is derived from redshift, and a further dependence on the RSD follows when the galaxy survey selection function is formulated in terms of redshift rather than true distance. Both of these conditions hold here. The effect of RSD is not large however \citet{Norberg_et_al}, and we do not consider it further.

We calculate $\xi (r_p, \upi )$ using the Landy \& Szalay estimator~\citep{Landy_Szalay},
\begin{equation}\label{eq:landy_szalay}
    \xi\left(r_{p}, \upi\right)
    =
    \frac{\mathrm{DD} - 2\, \mathrm{DR} + \mathrm{RR}}{\mathrm{RR}},
\end{equation}
with $\mathrm{DD}$, $\mathrm{DR}$, and $\mathrm{RR}$ being the numbers of (possibly weighted) data-data, data-random and random-random pairs normalised by the total number of pairs. $\xi$ is related to the excess probability of finding a galaxy pair separated by the redshift space distance $\sqrt{\upi^2 + r_p^2}$, relative to a uniform distribution of galaxies. The excess probability can be written as 
\begin{equation}\label{eq:excess_probability}
    \mathrm{d}P
    =
    \bar{n}^2
    \left[ 1 + \xi\left(r_p, \upi\right) \right]\mathrm{d}V_1 \mathrm{d}V_2,
\end{equation}
where $\bar{n}$ is the mean galaxy number density and $\mathrm{d}V_{1, 2}$ is the volume element associated with the two galaxies. For a uniform distribution $\xi = 0$, $\xi > 0$ indicates clustering of galaxies, and $\xi < 0$ means galaxies are less clustered than a uniform distribution.

When calculating $\xi$ via Eq.~\eqref{eq:landy_szalay} one of the required inputs is a distribution of uniformly distributed galaxies spanning the same volume as the observed galaxies. To ensure the same angular distribution, we use the \ac{NYU} large scale structure sample's random catalogue, which also matches the \ac{NSA} angular geometry. For the matched catalogue, we draw uniformly distributed samples on the sky from within the ``mangle'' polygons that define the survey geometry. Mangle is a software designed to deal with complex angular masks~\citep{Mangle}. To each of these randomly-drawn points we assign the galaxy properties of interest (e.g. absolute magnitude and redshift) by randomly drawing samples from the survey, and then apply the desired absolute magnitude cut. This method has been verified for wide-angle surveys and avoids the need to explicitly model the surveys' radial survival function~\citep{Ross_et_al}, while providing a unified method for our optically-selected and $\HI$-selected catalogues.

\begin{figure*}
    \centering
    \includegraphics[width=1.0\textwidth]{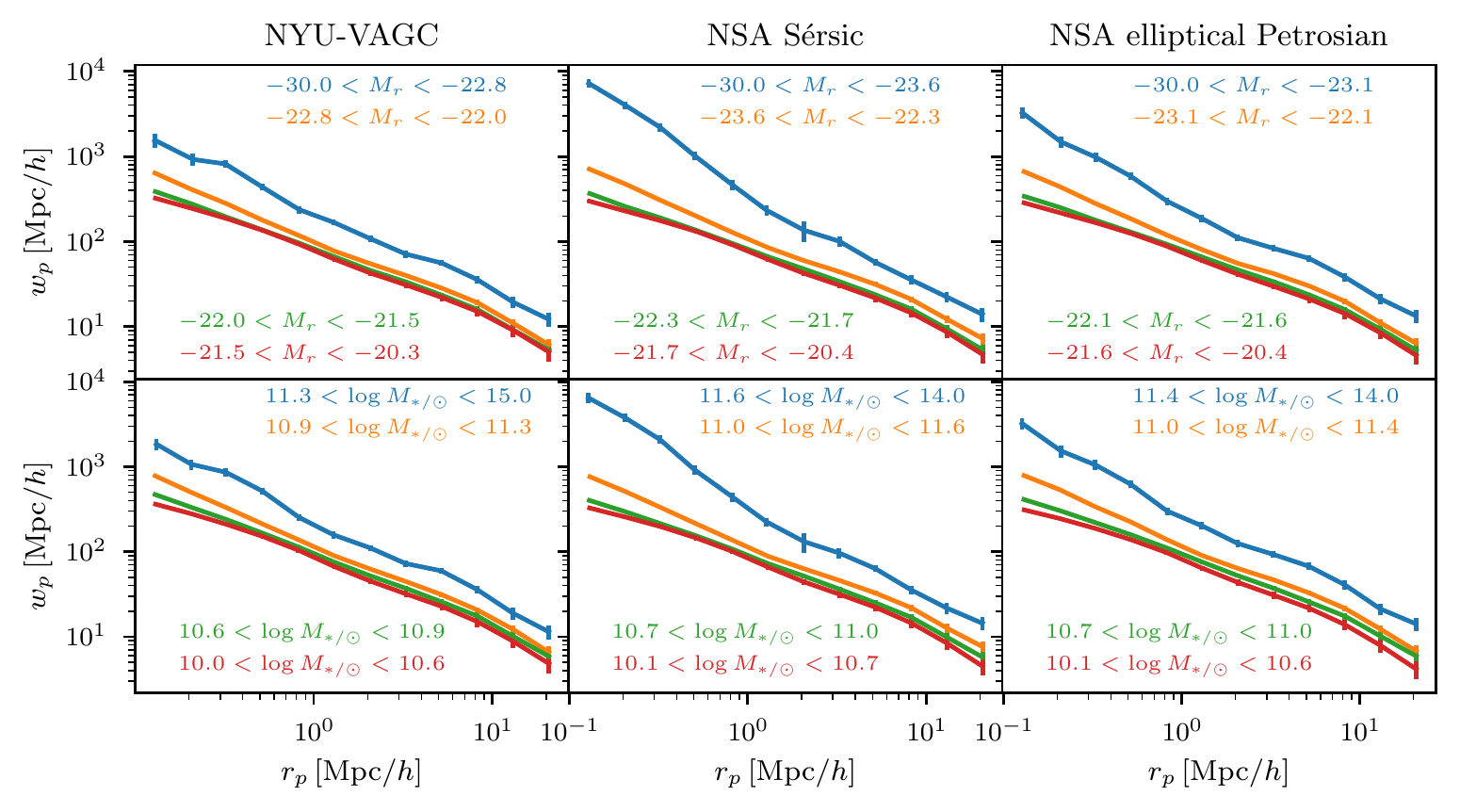}
    \caption{Two-point projected correlation function $w_p$ for \ac{NYU} Petrosian, \ac{NSA} S\'ersic and \ac{NSA} elliptical Petrosian photometries. Subsamples in the top row are defined by $r$-band luminosity while those in the bottom row are defined by stellar mass. These are the correlation functions used to calibrate our \ac{SHAM} models. In particular, the small-scale two-point correlation function is sensitive to the presence of satellite galaxies which boost the clustering. The percentile selection used to determine the cuts in target quantities is given in Table~\ref{tab:galaxy_counts}.
    In what follows, the brightest (most massive) cut will be referred to as ``1st subsample'', the next cut as ``2nd subsample'' and so on. We denote $M_* / M_\odot$ by $M_{*/\odot}$.}
    \label{fig:optical_clustering}
\end{figure*}

\begin{figure*}
    \centering
    \includegraphics[width=1.0\textwidth]{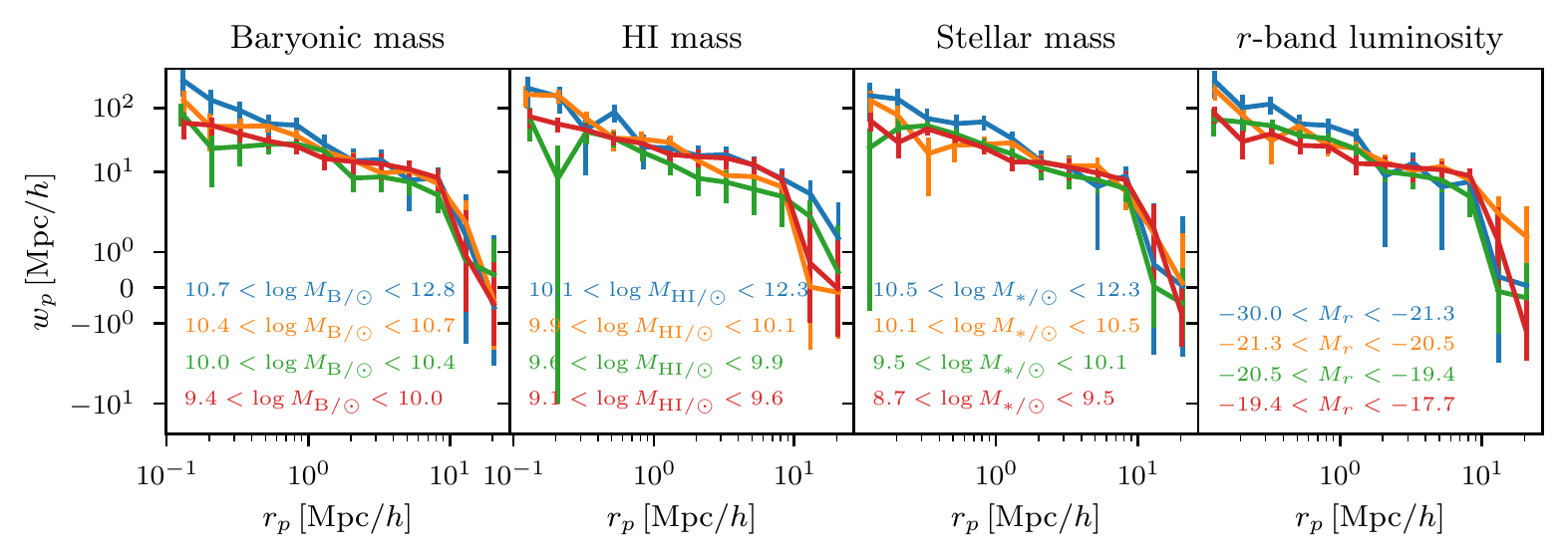}
    \caption{Two-point projected correlation function for the \matched catalogue. The panels are distinguished by the galaxy property used to define the subsamples. The percentile selection used to determine the cuts in target quantities is given in Table~\ref{tab:galaxy_counts_matched}. We define $M_{x / \odot} \equiv M_x / M_\odot$, where $M_x$ is either the baryonic, $\HI$, or stellar mass.}
    \label{fig:HI_clustering}
\end{figure*}

The uncertainty on the survey's two-point projected space correlation function $w_p$ is estimated via the jackknife resampling method~\citep{Norberg_et_al}. We split the survey into 256 clusters in RA-dec by using the \textsc{kmeans-radec} algorithm.\footnote{\url{https://github.com/esheldon/kmeans_radec}} These clusters are calculated using the uniformly spaced random samples, to which the survey's galaxies are then assigned. We calculate the correlation function, with the $k$-th cluster excluded, $\bm{w}_{p_k}$; the jackknife covariance matrix is then calculated as
\begin{equation}\label{eq:jackknife}
    \bm{\mathrm{C}}_{\mathrm{jack}}
    =
    \frac{N - 1}{N}
    \sum_{k=1}^{N}
    \left(\bm{w}_{p_k} - \widehat{\bm{w}}_{p}\right)
    \left(\bm{w}_{p_k} - \widehat{\bm{w}}_{p}\right)^\intercal
\end{equation}
where $\widehat{\bm{w}}_{p}$ denotes the mean vector of all $N$ jackknife estimates and $N$ is the total number of clusters.
We show the galaxy survey correlation functions for the \ac{NYU} and \ac{NSA} catalogues in Fig.~\ref{fig:optical_clustering}, where we bin the galaxies by $M_r$ or $\log M_*/M_\odot$. Similarly, Fig.~\ref{fig:HI_clustering} displays the galaxy correlation functions for the matched catalogue, where the galaxies are binned either by $\log M_\mathrm{B}/M_\odot$, $\log M_{\HI}/M_\odot$, $\log M_*/M_\odot$ or $M_r$.

We also estimate the uncertainty on the \ac{SHAM} mocks' two-point correlation function via jackknifing. However, as the simulated galaxies are distributed over a periodic box, here we leave out one sub-volume of size $25 \times 25 \times 400~\mathrm{Mpc}^{3}/h^3$ at a time. Furthermore, as we introduce scatter into the galaxy--halo relation through the \ac{SHAM} parameter $\scatter$, at each point in the \ac{SHAM} parameter space we generate $50$ independent mocks to estimate the ``stochastic'' contribution to the covariance matrix.

\begin{figure*}
    \centering
    \includegraphics[width=1.0\textwidth]{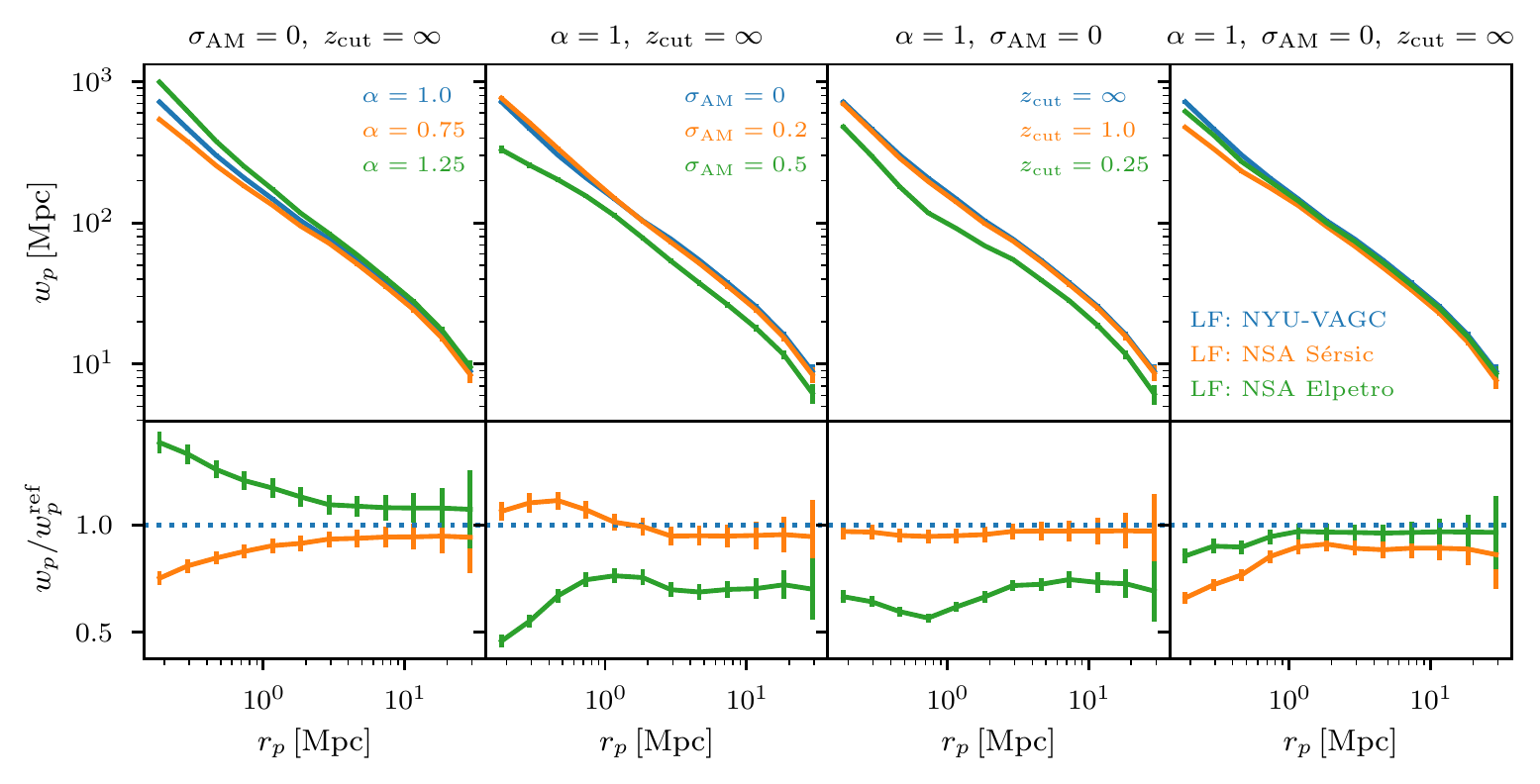}
    \caption{Variation of the simulated projected two-point correlation function $w_p$ for \ac{NYU} \ac{LF}-based \ac{SHAM} with the model parameters $\alpha$, $\scatter$, and $\zcut$ (see Sec.~\ref{sec:catalogs}). The fourth panel shows $w_p$ at a fixed posterior point but varying the input \ac{LF}. The panel titles indicate the parameters that are kept fixed.} 
    \label{fig:wps_vary_parameters}
\end{figure*}

In Fig.~\ref{fig:wps_vary_parameters} we show how varying the \acl{SHAM} parameters affects the two-point correlation function of the simulated catalogues (introduced in Sec.~\ref{sec:CF}). In particular, one by one we vary $\alpha$, $\scatter$, and $\zcut$ while keeping the other \ac{SHAM} parameters fixed in the first three panels of Fig.~\ref{fig:wps_vary_parameters}. The residuals are always taken with respect to the first value in each panel's legend. Increasing $\alpha$ gives stronger clustering on all scales as it results in preferentially ranking haloes with $M_{\mathrm{peak}} > M_0$. Increasing the scatter $\scatter$ in the galaxy-halo connection decreases clustering across all scales as the up-scattered galaxies dominate over the down-scatter galaxies. However, it is possible that initially for small values of $\scatter$ the up-scattering dominates and boosts clustering on small scales. Lastly, introducing the $\zcut$ threshold results in lowered clustering, as the earlier-forming haloes tend to cluster more strongly.

\subsection{Likelihood Framework}\label{sec:likelihood}

We split each catalogue into several lower and upper-limited bins. By using bins with no objects in common, as opposed to thresholds as is more common, we {minimise} covariance between subsamples when combining their separate posteriors below. We use uniform priors and define a Gaussian likelihood over the \ac{SHAM} two-point projected correlation functions, centred at the survey correlation function. The covariance matrix is taken to be the sum of the survey jackknife covariance, the \ac{SHAM} jackknife and the \ac{SHAM} covariance matrices. Therefore, the likelihood of the correlation function calculated from the \ac{SHAM} simulation $\bm{x} = \bm{w}_{p}^{\mathrm{AM}}\left(\bm{\theta}\right)$ as a function of the model parameters (typically $\alpha$ and $\scatter$) is
\begin{equation}
    p\left(\D \mid \bm{\theta}\right)
    =
    \frac{1}{\sqrt{\left|2 \pi \bm{\Sigma}\right|}}
    \exp{\left[-\frac{1}{2} \left(\bm{x} - \bm{\mu}\right)^\intercal
    \bm{\Sigma}^{-1}
    \left(\bm{x} - \bm{\mu}\right)
    \right]},
\end{equation}
where $\bm{\Sigma}$ is the aforementioned sum of covariance matrices, $\bm{\mu}$ is the survey correlation function, and $k$ is the number of $r_p$ bins.

We compare the goodness-of-fit of \ac{SHAM} in various cases by means of the Bayesian evidence. For a single catalogue, with  $\bm{\theta}_i$ and $\D_i$ the fitted parameters and data in the $i$-th bin respectively, the evidence is given by
\begin{equation}
    p\left(\D_i \mid I\right)
    =
    \int p\left(\D_i \mid \bm{\theta}_i, I\right)
    p\left(\bm{\theta}_i \mid I \right) \mathrm{d}^{n}\bm{\theta},
\end{equation}
where $I$ denotes the model under consideration, which in our case reflects the choice of galaxy and halo proxy in \ac{SHAM}, and the photometry and selection criteria of the sample. $p\left(\D_i \mid \bm{\theta}_i, I\right)$ is the likelihood of the data for given model parameters, while $p\left(\bm{\theta}_i \mid I \right)$ is the prior on the parameters. To quantify parameter tension between bins we can compare the evidence of a model in which all bins share the same parameters to one in which the parameters are unique to each bin:
\begin{equation}\label{eq:bayes_tension}
    \mathcal{B}_\mathrm{tension}
    = 
    \frac
    {\int \Big[ \prod_i p\left(\D_i \mid \bm{\theta}, I\right)\Big]
    p\left(\bm{\theta} \mid I \right) \mathrm{d}^{n}\bm{\theta}
    }
    {\prod_i \Big[ \int p\left(\D_i \mid \bm{\theta}_i, I\right)
    p\left(\bm{\theta}_i \mid I \right) \mathrm{d}^{n}\bm{\theta}_i \Big]
    },
\end{equation}
where the product over $i$ covers the bins that we wish to compare~\citep{Evidence_test}. This yields the Bayes factor quantifying whether all bins share the same posterior parameters $\bm{\theta}$, or if each bin is better described by different parameters $\bm{\theta}_i$. This will allow us to answer the question of whether there is evidence for the running of \ac{SHAM} parameters as a function of galaxy mass or luminosity.

Similarly, we can ask whether \ac{SHAM} is more accurate when performed with luminosity $M_r$ or stellar mass $M_*$. The Bayes factor for this comparison is
\begin{equation}\label{eq:bayes_proxy}
    \mathcal{B}_{\mathrm{proxy}}
    =
    \frac
    {p\left(\D \mid I=M_r\right)}
    {p\left(\D \mid I=M_*\right)}.
\end{equation}
Here we can either compare evidences for specific bins, or we can combine the constraints from several bins before evaluating Eq.~\ref{eq:bayes_proxy}. For the latter, the combined evidence can either be calculated as in the numerator or denominator of Eq.~\eqref{eq:bayes_tension}, according to whether or not we assume the \ac{SHAM} parameters are common between the bins. Lastly, we also calculate which photometric reduction pipeline is best-suited for \ac{SHAM}, which is analogous to Eq.~\eqref{eq:bayes_proxy} but where the model instead corresponds to \ac{NYU}, \ac{NSA} S\'ersic, or \ac{NSA} elliptical Petrosian photometry.

\section{Results}\label{sec:results}

\subsection{Optically-selected samples}

We constrain the \ac{SHAM} parameters $\alpha$ and $\scatter$ (Eq.~\ref{eq:halo_proxy}) by comparing the two-point projected correlation function from the \ac{SHAM} mocks to that of the observational data. As described above, by splitting each catalogue into subsamples defined by lower and upper thresholds of luminosity or mass we can address whether the halo proxy shows any dependence on these quantities, which of them is best-suited for \ac{SHAM}, and how the best-fit parameters respond to the photometric reduction.

\begin{table*}
\setlength{\tabcolsep}{3pt}
\renewcommand{\arraystretch}{1.25}
\begin{tabular}{llccccccccccc}
                            &   & \multicolumn{3}{c}{NYU}                         &  & \multicolumn{3}{c}{NSA S\'ersic}                  &  & \multicolumn{3}{c}{NSA Elpetro}                 \\ \cline{3-5} \cline{7-9} \cline{11-13} 
Percentile range            &   & $N$       & $M_r$      & $\log M_*/M_\odot$ &  & $N$       & $M_r$      & $\log M_*/M_\odot$ &  & $N$       & $M_r$      & $\log M_*/M_\odot$ \\ \cline{1-1} \cline{3-5} \cline{7-9} \cline{11-13} 
$98.5\% \rightarrow 100\%$ & & $8,040$   & $(-30.0, -22.8)$ & $(11.3, 15.0)$   &  & $7,407$   & $(-30, -23.6)$ & $(11.6, 15.0)$   &  & $7,090$   & $(-30.0, -23.1)$ & $(11.4, 15.0)$  \\
$85.0\% \rightarrow 98.5\%$  & & $72,353$  & $(-22.8, -22.0)$ & $(10.9, 11.3)$   &  & $66,695$  & $(-23.6, -22.3)$ & $(11.0, 11.6)$   &  & $63,810$   & $(-23.1, -22.1)$ & $(11.0, 11.4)$   \\
$60.0\% \rightarrow 85.0\%$  & & $133,988$ & $(-22.0, -21.5)$ & $(10.6, 10.9)$   &  & $123,508$ & $(-22.3, -21.7)$ & $(10.7, 11.0)$   &  & $118,166$   & $(-22.1, -21.5)$ & $(10.7, 11.0)$   \\
$20.0\% \rightarrow 60.0\%$  & & $214,381$ & $(-21.5, -20.3)$ & $(10.0, 10.6)$    &  & $197,615$ & $(-21.7, -20.4)$ & $(10.1, 10.7)$    &  & $189,065$  & $(-21.5, -20.4)$ & $(10.1, 10.7)$   \\ \cline{1-1} \cline{3-5} \cline{7-9} \cline{11-13}
\end{tabular}
\caption{\label{tab:galaxy_counts}Definition of the subsamples for the \ac{NYU} and \ac{NSA} samples. The subsamples are split by $M_r$ or $M_*$, with $100\%$ corresponding to the brightest galaxy in the parent sample. The two-point projected correlation functions corresponding to these subsamples are shown in Fig.~\ref{fig:optical_clustering}.}
\end{table*}

After applying the cuts described in Sec.~\ref{sec:data_obs}, we calculate the \acp{LF} and \acp{SMF} (Fig.~\ref{fig:LFs_MFs}) from the remaining galaxies, and the two-point projected correlation function for each catalogue subsample. The $1$st subsample is assigned the brightest (or most massive) $1.5\%$ of galaxies, the $2$nd subsample is assigned the next $13.5\%$, the $3$rd is assigned the next $25\%$, and finally the $4$th bin is assigned the next $40\%$ of galaxies. The faintest $20\%$ of the galaxies in each catalogue are excluded, since they may be incomplete and/or the corresponding simulated haloes may not be well-resolved. This binning scheme allows for a selection of a comparable number of galaxies in each subsample despite the different \ac{SDSS} photometric reductions used, and the same number of objects when comparing $M_r$ to $M_*$. The samples do not share any objects, minimising their cross-correlation and statistical dependence. The binning is summarised in Table~\ref{tab:galaxy_counts} for all optically-selected samples that we consider (\ac{NYU} Petrosian, \ac{NSA} S\'ersic, and \ac{NSA} elliptical Petrosian). In what follows we will refer to the brightest (or most massive) subsample as the ``$1$st subsample'', second brightest as ``$2$nd subsample'', third brightest as ``$3$rd subsample'' and faintest as ``$4$th subsample''.

\begin{figure*}
    \centering
    \includegraphics[width=\textwidth]{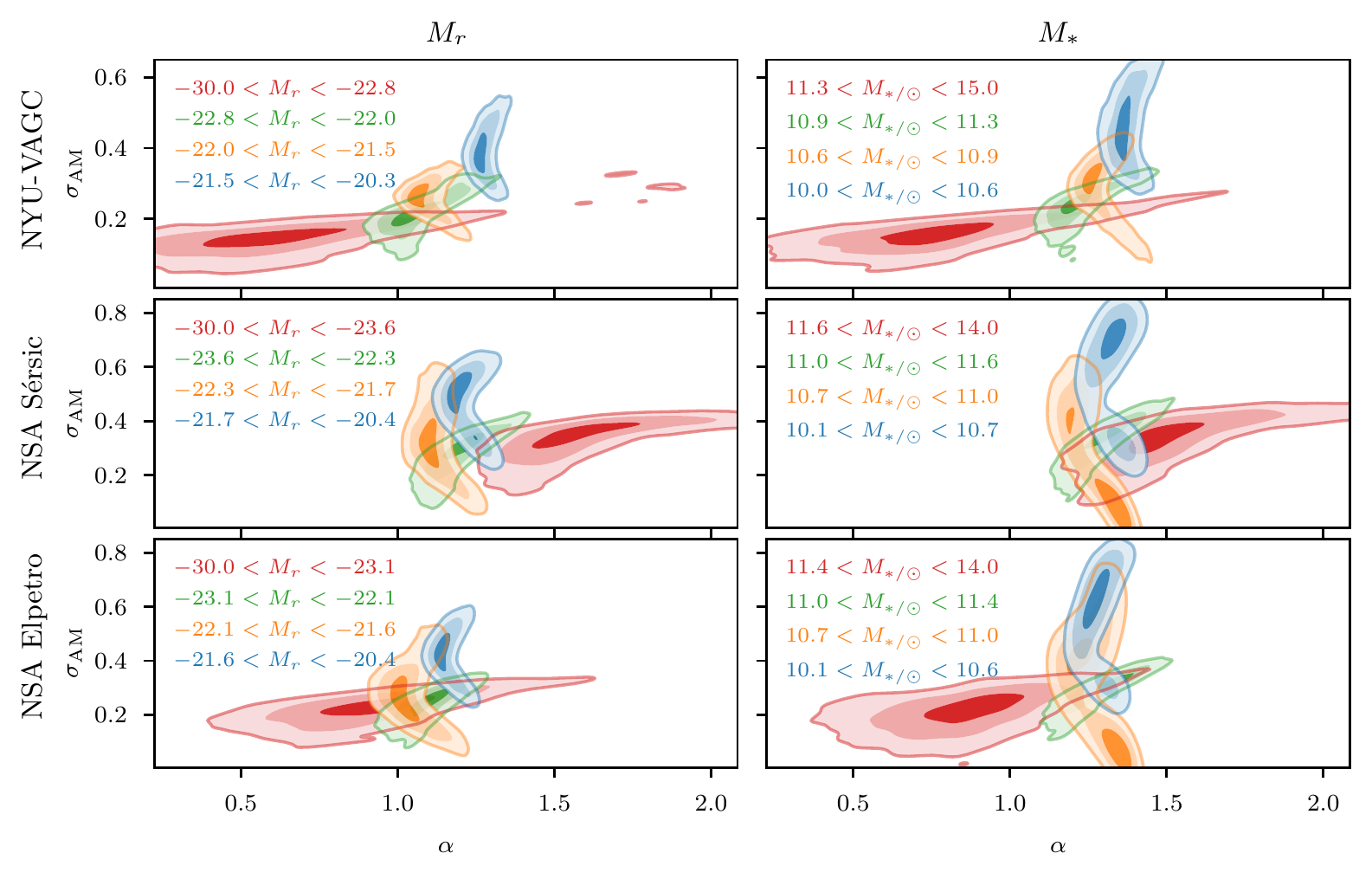}
    \caption{Posteriors on $\alpha$ and $\scatter$ for optically-selected samples. The contours show the minimal areas enclosing $39\%$, $86\%$ and $99\%$ of the sample. The rows distinguish different catalogues, while columns correspond to $M_r$ and $M_*$. Each panel shows the posteriors for each of the four subsamples, from brightest in red to faintest in blue.}
    \label{fig:posterior_contours}
\end{figure*}

\begin{figure*}
    \centering
    \includegraphics[width=1.0\textwidth]{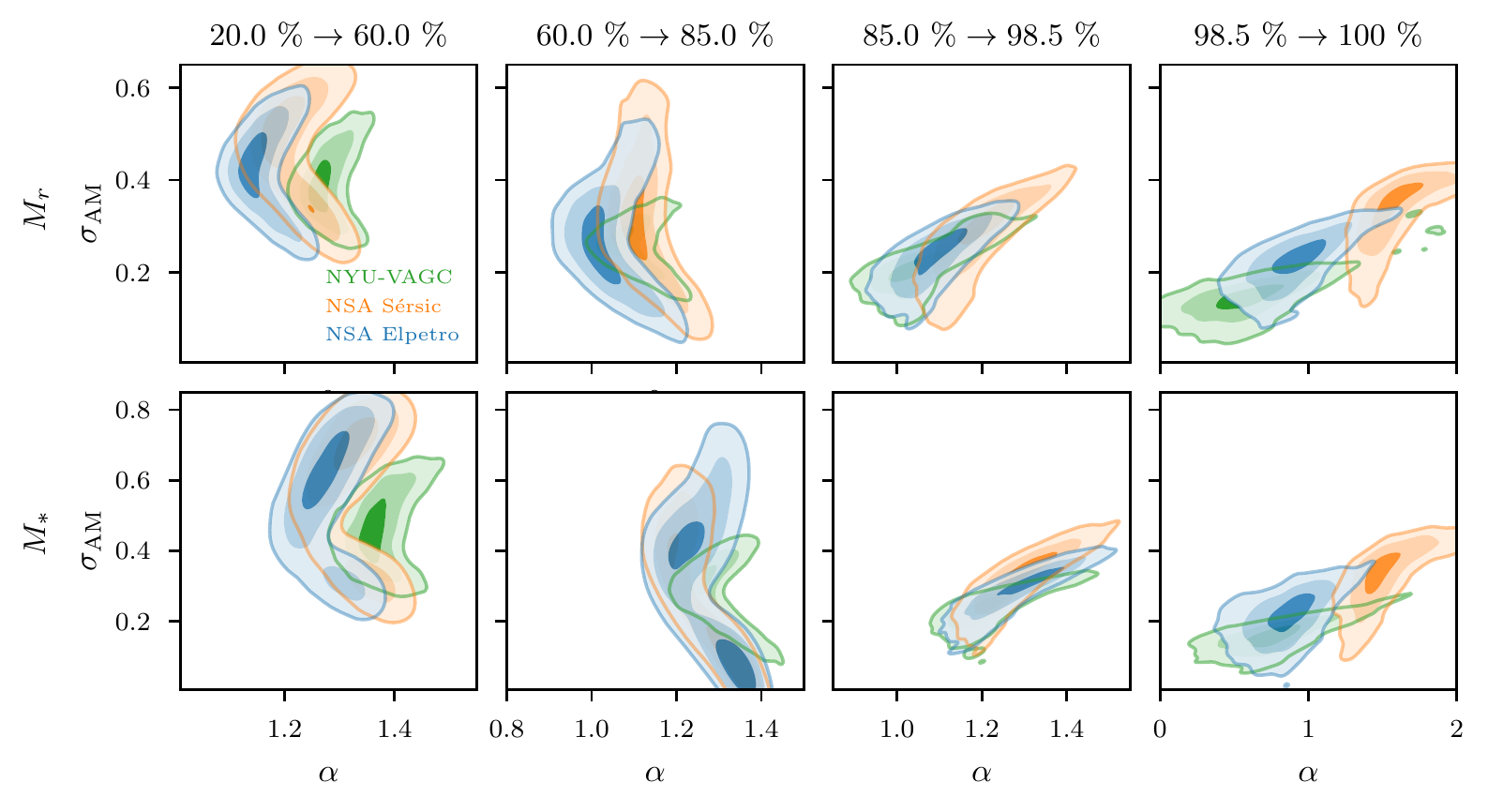}
    \caption{Direct comparison of posteriors on $\alpha$ and $\scatter$ ($39\%$, $86\%$ and $99\%$ levels) between the \ac{NYU} Petrosian, \ac{NSA} S\'ersic and \ac{NSA} elliptical Petrosian samples. Rows distinguish $M_r$ vs and $M_*$-based \ac{SHAM} while columns distinguish the subsamples. Each subsample contains galaxies between the two percentile ranges indicated in the title, where $100\%$ corresponds to the brightest or most massive galaxy. The subsamples are defined in Table~\ref{tab:galaxy_counts}.}
    \label{fig:comparison_nsa2}
\end{figure*}

The posteriors on the \ac{SHAM} parameters for each subsample are shown in Fig.~\ref{fig:posterior_contours}, where the rows reflect the three samples. Typically both $M_r$ and $M_*$ display similar posterior shapes in the $\alpha-\scatter$ plane with the degeneracy between $\alpha$ and $\scatter$ and the dependence of the maximum-likelihood parameters on $M_r$ and $M_*$ being clearly visible. Some degeneracy between $\alpha$ and $\scatter$ should be expected because both control the simulated galaxy clustering, as shown in Fig.~\ref{fig:wps_vary_parameters}. Increasing $\alpha$ boosts the fraction of subhaloes matched to our galaxies (Fig.~\ref{fig:subhalo_fractions}), and consequently increases the small-scale clustering. On the other hand, increasing $\scatter$ weakens the clustering signal. In each sample we typically find that the fainter (or less massive) subsamples prefer larger values of $\scatter$, and that $M_*$-based \ac{SHAM} requires larger $\alpha$ and $\scatter$ than the $M_r$-based model. For a direct comparison of the samples, we show Fig.~\ref{fig:comparison_nsa2} where the individual plots display the comparison of \ac{SHAM} model parameters between \ac{NYU} and \ac{NSA} catalogues. We find that all catalogues show a similar \ac{SHAM} parameter dependence in the $2$nd and $3$rd subsample. In the $4$th (faintest) subsample a marginal difference between \ac{NYU} and \ac{NSA} catalogues is visible, with the \ac{NYU} subsamples preferring larger values of $\alpha$. However, most notably, in the $1$st subsample (brightest and most massive) \ac{NSA} S\'ersic shows substantially stronger preference for larger values of alpha than the two Petrosian photometries, and therefore for populating stripped haloes. The fainter subsamples show a distinct ``banana shaped'' posterior, which follows from our use of lower and upper bounds to define the samples. At relatively low values of $\scatter$, the down-scattering of massive haloes can actually boost clustering, although for higher scatter values the up-scattering of less massive haloes dominates and lowers the clustering.

\begin{figure*}
    \centering
    \includegraphics[width=1.0\textwidth]{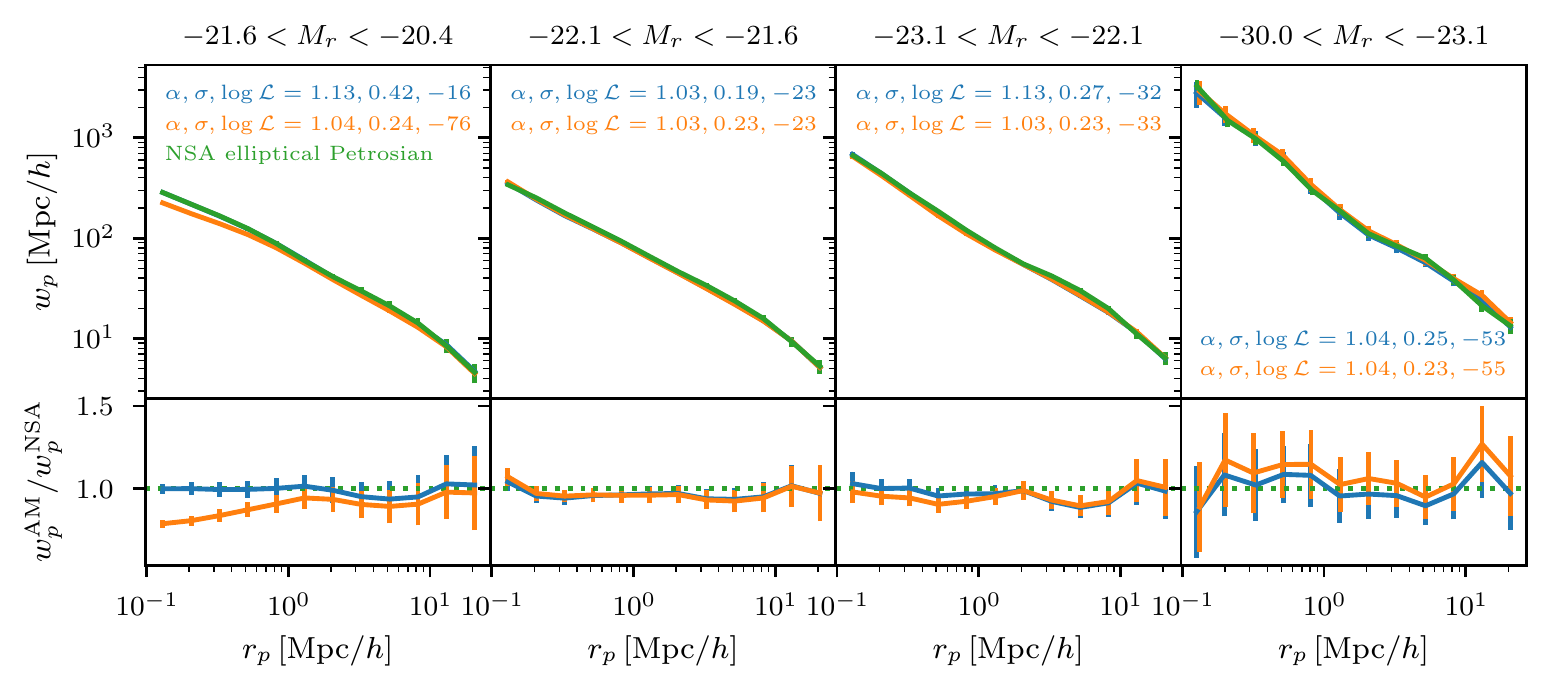}
    \caption{Comparison of each sample's maximum-likelihood \ac{SHAM} two-point projected correlation function and the maximum-likelihood combined constraint from the three brightest bins to the survey correlation function. Here we show the \ac{NSA} elliptical Petrosian luminosity-based \ac{SHAM}. $\log\mathcal{L}$ denotes maximum log-likelihood values and the residuals are taken with respect to the survey's $w_p$.}
    \label{fig:fits_NSA_elpetro_LF}
\end{figure*}

\begin{figure}
    \centering
    \includegraphics[width=1.0\columnwidth]{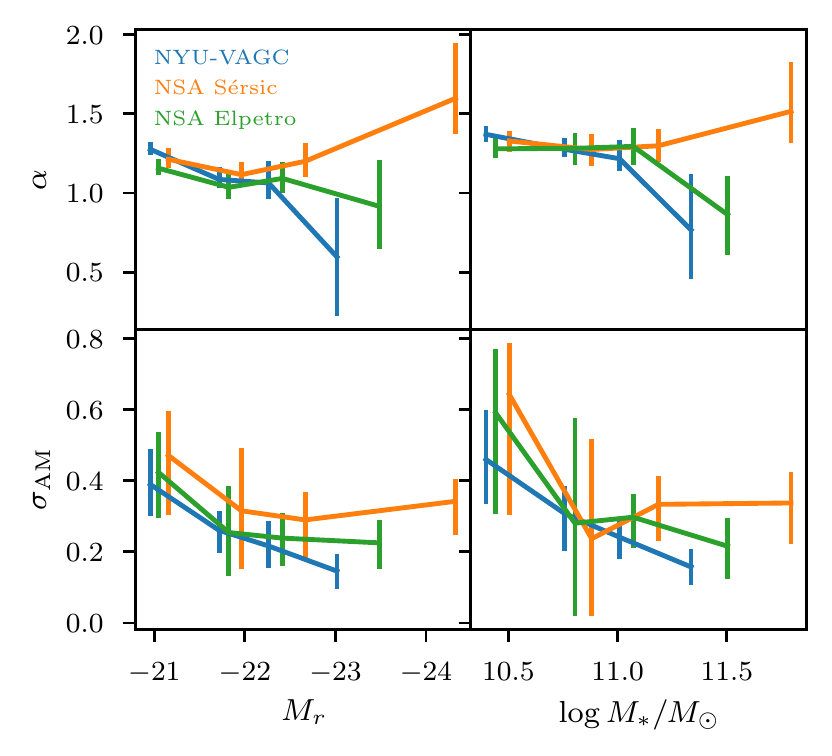}
    \caption{Best-fit \ac{SHAM} parameters, and their $90\%$ uncertainties, for our optically-selected samples as a function of $M_r$ and $M_*$. The scatter in the galaxy--halo connection increases towards the faint end, while variation in the halo proxy is weak and sample-dependent.}
    \label{fig:LF_SMF_dependence}
\end{figure}

To answer the question of whether the \ac{SHAM} galaxy--halo connection model shows systematic variation with galaxy brightness or mass, we analyse the individual posteriors following the formalism of Sec.~\ref{sec:likelihood} by comparing the models' evidences. The Bayes factors are summarised in Table~\ref{tab:parameter_tension}. We quote the tensions $\mathcal{B}_{12}$, $\mathcal{B}_{23}$, $\mathcal{B}_{34}$, $\mathcal{B}_{14}$, $\mathcal{B}_{123}$, and $\mathcal{B}_{1234}$, where the subscript $ij$ gives the tension between $i$-th and $j$-th bin. No statistically significant evidence for either luminosity or stellar mass dependence is observed over the three brightest or most massive subsamples in either photometry sample. However, the \ac{NYU} Petrosian sample shows a strong tension between the 1st and 4th subsample in both $M_r$ and $M_*$. Weaker tension is also observed in the other samples, indicating that the variation in the \ac{SHAM} parameters occurs over a larger range of brightness and mass for \ac{NSA} S\'ersic and elliptical Petrosian samples. Consequently, we conclude that all four subsamples are mutually incompatible in all samples (and in both $M_r$ and $M_r$) under the assumption of universal SHAM parameters, a result usually driven by the faintest subsample. As there is no statistically significant luminosity or stellar-mass dependence over the first three subsamples (except for the \ac{NSA} S\'ersic $M_r$ sample), we quote the $90\%$ confidence intervals on $\alpha$ and $\scatter$ in Table~\ref{tab:combined_posteriors_results}. As an example, the best two-point projected correlation function fits for the \ac{NSA} elliptical Petrosian luminosity sample, along with the combined posterior fit, are shown in Fig.~\ref{fig:fits_NSA_elpetro_LF}. We summarise how the best-fit values of $\alpha$ and $\scatter$ (and their $90\%$ confidence intervals) depend on the subsample index -- which can be viewed as a discrete sampling in $M_r$ or $M_*$ -- in Fig.~\ref{fig:LF_SMF_dependence}.

\begin{table}
\setlength{\tabcolsep}{2.25pt}
\renewcommand{\arraystretch}{1.25}
\begin{tabular}{lcccccc}
 & $\mathcal{B}_{12}$ & $\mathcal{B}_{23}$ & $\mathcal{B}_{34}$ & $\mathcal{B}_{14}$ & $\mathcal{B}_{123}$ & $\mathcal{B}_{1234}$ \\
 \hline
\ac{NYU} $M_r$          & $3.2$ & $17$  & \nstrong $0.029$ & \Nstrong $2.1\times 10^{-5}$ & $3.7$ & \Nstrong $9.4 \times 10^{-11}$ \\
\ac{NYU} $M_*$          & $2.6$ & $18$  & $3.5$   & \Nstrong $8.0\times 10^{-5}$ & $0.60$ & \Nstrong $1.4\times 10^{-11}$ \\
\ac{NSA} S\'ersic $M_r$   & $0.060$ & $5.0$ & $0.74$ & $0.10$ & \nstrong $0.028$ & \Nstrong $4.2\times 10^{-5}$ \\
\ac{NSA} S\'ersic $M_*$   & $2.4$ & $1.7$ & $0.33$ & $1.2$ & $1.09$ & \Nstrong $1.6 \times 10^{-3}$ \\
\ac{NSA} Elpetro $M_r$  & $10$ & $14$   & $0.67$ &  $0.59$ & \Pstrong $190$ & \Nstrong $5.6\times 10^{-6}$ \\
\ac{NSA} Elpetro $M_*$  & $0.50$ & $0.39$   & $3.2$ & \nstrong $0.050$ & $0.23$ &  \Nstrong $2.4\times 10^{-3}$\\
\hline
\end{tabular}
\caption{\label{tab:parameter_tension}
Bayes factors $\mathcal{B}_i$ quantifying the consistency between $\alpha$ and $\scatter$ among various bin combinations $i$. $\mathcal{B} > 1$ indicate no tension, i.e. no evidence for dependence on $M_r$ or $M_*$, while $\mathcal{B} \ll 1$ indicates significant tension. Faint green is used to distinguish strong support ($20 \leq \mathcal{B}_i < 150$) for consistency, deep green denotes very strong support ($150 \leq \mathcal{B}$) for consistency, and correspondingly in shades of red for tension. We find that the faintest (4th) bin shows strong evidence for different \ac{SHAM} parameters to the other subsamples.}
\end{table}

\begin{table}
\renewcommand{\arraystretch}{1.25}
\begin{tabular}{lccc}
                        & $\alpha$               & $\scatter/\mathrm{dex}$               & Domain                  \\ \hline
\ac{NYU} $M_r$         & $1.09^{+0.03}_{-0.03}$ & $0.21^{+0.01}_{-0.01}$ & $M_r < - 21.5$           \\
\ac{NYU} $M_*$    & $1.25^{+0.03}_{-0.03}$ & $0.24^{+0.02}_{-0.02}$ & $M_* > 10^{10.6}M_\odot$        \\
\ac{NSA} S\'ersic $M_r$  & $1.15^{+0.03}_{-0.03}$ & $0.24^{+0.05}_{-0.05}$ & $M_r  < - 21.7$          \\
\ac{NSA} S\'ersic $M_*$ & $1.23^{+0.03}_{-0.03}$ & $0.26^{+0.05}_{-0.05}$ & $M_*> 10^{10.7}M_\odot$     \\
\ac{NSA} Elpetro $M_r$ & $1.03^{+0.04}_{-0.04}$ & $0.23^{+0.03}_{-0.03}$ & $M_r< -21.5$             \\
\ac{NSA} Elpetro $M_*$  & $1.20^{+0.03}_{-0.03}$ & $0.27^{+0.03}_{-0.03}$ & $M_* > 10^{10.7} M_\odot$  \\
\hline
\end{tabular}
\caption{\label{tab:combined_posteriors_results} Maximum-likelihood and $90\%$ credible intervals of $\alpha$ and $\scatter$ from the first three bins in luminosity or stellar mass (see Table~\ref{tab:galaxy_counts}).}
\end{table}

We can also compare the goodness-of-fit between $M_r$- and $M_*$-based \ac{SHAM} models. First, we compare the evidences from the individual subsamples, as shown in the first four columns of Table~\ref{tab:Mr2M*_comparison}. We find a weak preference for the luminosity-based model in each case. Next, if we assume that the halo proxy is specific to each subsample (``local'' subscript in Table~\ref{tab:Mr2M*_comparison}), this yields strong preference for the luminosity-based model in each sample. Lastly, we can also combine the evidence from the first three subsamples (found not to be mutually exclusive) assuming the halo proxy not to vary across the subsamples (``global'' subscript in Table~\ref{tab:Mr2M*_comparison}), producing again strong evidence in favour of $M_r$ in the two Petrosian samples and inconclusive evidence in case of the \ac{NSA} S\'ersic sample. We therefore conclude that for Petrosian photometry, luminosity-based \ac{SHAM} reconstructs observed the clustering significantly better than stellar mass-based \ac{SHAM} across the range of galaxy brightness that we consider.

\begin{table}
\renewcommand{\arraystretch}{1.25}
\setlength{\tabcolsep}{5pt}
\begin{tabular}{lccccccc}
 &
  $\mathcal{B}_1$ &
  $\mathcal{B}_2$ &
  $\mathcal{B}_3$ &
  $\mathcal{B}_4$ &
  $\mathcal{B}_{123}^{\mathrm{local}}$ &
  $\mathcal{B}_{123}^{\mathrm{global}}$ &
  $\mathcal{B}_{1234}^{\mathrm{local}}$ \\ \hline
NYU         & $2.5$ & $4.8$  & $2.2$ & $3.4$ & $9.5$ & \pstrong $140$ & $13$ \\
NSA S\'ersic  & $0.64$ & $16$ & $7.9$ & $2.4$ & \pstrong $81$   & $1.7$  & \Pstrong $190$    \\
NSA Elpetro & $2.7$ & $3.44$  &  $15$ & $15$  & \pstrong $140$   & \Pstrong $9.5 \times 10^4$ & \Pstrong $2200$   \\
\hline
\end{tabular}
\caption{\label{tab:Mr2M*_comparison} Bayes factors $\mathcal{B}_i$ quantifying the preference for $M_r$ over $M_*$-based \ac{SHAM} in bin(s) $i$. $\mathcal{B}>1$ indicates preference for the $M_r$-based sample. The ``global'' and ``local'' superscripts indicate whether we assume the halo proxy to be common to all subsamples (global), or peculiar to each subsample (local). The colours follow the convention of Table~\ref{tab:parameter_tension}.}
\end{table}

Having explored each catalogue, we now address the question of which catalogue's clustering can be best fitted with \ac{SHAM}. In luminosity, considering only the first three, mutually consistent subsamples, we find the strongest preference for the \ac{NYU} Petrosian photometry, with strong support relative to \ac{NSA} elliptical Petrosian photometry ($\mathcal{B} = 43 : 1$), and very strong support compared to \ac{NSA} S\'ersic ($\mathcal{B} = 5.3 \times 10^{6} : 1$). The other possible comparison is across all four bins with luminosity dependence of the parameters: this yields again a very strong support for \ac{NYU} photometry relative to both \ac{NSA} elliptical Petrosian ($\mathcal{B} = 190 : 1$) and S\'ersic photometry ($\mathcal{B} = 6000 : 1$). We therefore conclude that the photometric reduction of \ac{NYU} -- as has been used in most previous studies -- is best suited for luminosity-based \ac{SHAM} modelling.

The catalogues can also be compared for stellar mass-based \ac{SHAM}. Over the first three, mutually consistent subsamples \ac{NYU} Petrosian photometry is strongly preferred over \ac{NSA} elliptical Petrosian ($\mathcal{B} = 2.4 \times 10^4 : 1$) and S\'ersic ($\mathcal{B} = 4.4\times 10^{4} : 1$) photometries. Finally, considering all four bins and assuming stellar mass dependence of the proxy yields again strong support for \ac{NYU} Petrosian photometry over \ac{NSA} elliptical Petrosian ($\mathcal{B} = 4600 : 1$) and S\'ersic photometry ($\mathcal{B} = 1.2\times 10^4 : 1$). \ac{NYU} Petrosian stellar masses are therefore the best for use in \ac{SHAM}.

\subsection{HI-selected samples}

\begin{table*}
\begin{tabular}{lccccc}
Percentile range             & $N$    & $M_r$         & $\log M_*/M_\odot$    & $\log M_\mathrm{B}/M_\odot$    & $\log M_{\HI}/M_\odot$  \\ \hline
$87.5\% \rightarrow 100\%$ & $1214$ & $(-24.8, -21.3)$ & $(10.5, 12.3)$ & $(10.8, 12.3)$ & $(10.1, 11.3)$ \\
$67.5\% \rightarrow 87.5\%$  & $1943$ & $(-21.3, -20.5)$ & $(10.1, 10.5)$  & $(10.4, 10.8)$ & $(9.9, 10.1)$  \\
$40.0\% \rightarrow 67.5\%$  & $2673$ & $(-20.5, -19.4)$ & $(9.5, 10.1)$   & $(10.0, 10.4)$  & $(9.6, 9.9)$  \\
$10.0\% \rightarrow 40.0\%$  & $2916$ & $(-19.4, -17.7)$ & $(8.7, 9.5)$    & $(9.4, 10.0)$   & $(9.1, 9.6)$    \\ \hline
\end{tabular}
\caption{\label{tab:galaxy_counts_matched} Definition of subsamples for the \matched catalogue.}
\end{table*}

For the $\HI$-selected samples we use our \matched catalogue (see Sec.~\ref{sec:data_obs}). We split the matched catalogue into $4$ independent samples by stellar, $\HI$ or baryonic mass, or $r$-band absolute magnitude, as summarised in Table.~\ref{tab:galaxy_counts_matched}. Because of the smaller survey volume compared to \ac{SDSS}, the \matched catalogue contains significantly fewer massive galaxies. Therefore, to ensure that our subsamples are not noise dominated, we alter our binning scheme: the $1$st subsample contains the top $12.5\%$, the $2$nd subsample the next $20\%$, the $3$rd subsample the next $27.5\%$ and the final $4$th subsample has the next $30\%$ of galaxies. We calculate the projected correlation function (see Fig.~\ref{fig:HI_clustering}) and luminosity and mass function (see Fig.~\ref{fig:LFs_MFs}) for each subsample following the formalism of Secs.~\ref{sec:CF}~and~\ref{sec:LF}.

\begin{figure}
    \centering
    \includegraphics[width=1.0\columnwidth]{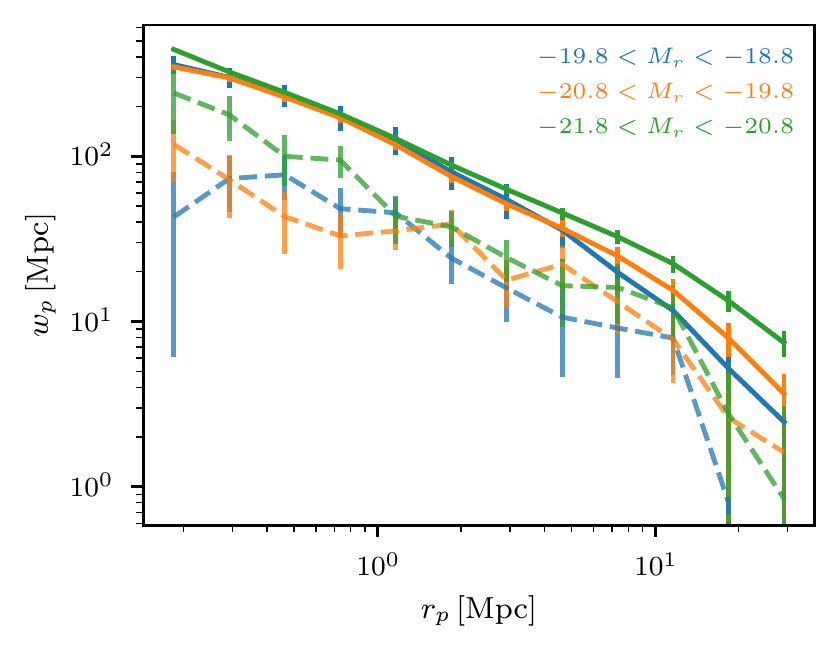}
    \caption{Comparison of optical (solid) vs $\HI$-selected (dashed) two-point projected correlation functions in bins of luminosity from the \ac{NSA} and \matched catalogues, respectively. Both the $\HI$- and optically-selected samples use \ac{NSA} elliptical Petrosian photometry, yet the former are significantly more weakly clustered.}
    \label{fig:HI2opt_clustering}
\end{figure}

Because of the $\HI$-selection, the samples are substantially less clustered than the optically-selected ones. This is shown in Fig.~\ref{fig:HI2opt_clustering}, where we compare the clustering of the \matched catalogue to the NSA elliptical Petrosian sample for different subsamples in $M_r$ (note that those subsamples are not the ones over which we perform \ac{SHAM} and serve only for illustrative purposes here). By studying the matched catalogue we will be able to determine how optical vs $\HI$ selection affects the results of \ac{SHAM} and whether \ac{SHAM} can model clustering of $\HI$-selected galaxies.
We begin with our \ac{SHAM} proxy of Eq.~\eqref{eq:halo_proxy}. As we will find that this model can only model $\HI$ clustering at very high $\scatter$, we then proceed to modify the \ac{SHAM} model to pre-select haloes with peak-mass redshift lower than $\zcut$, and then ranking haloes by present-day virial mass.

\begin{figure*}
    \centering
    \includegraphics[width=1.0\textwidth]{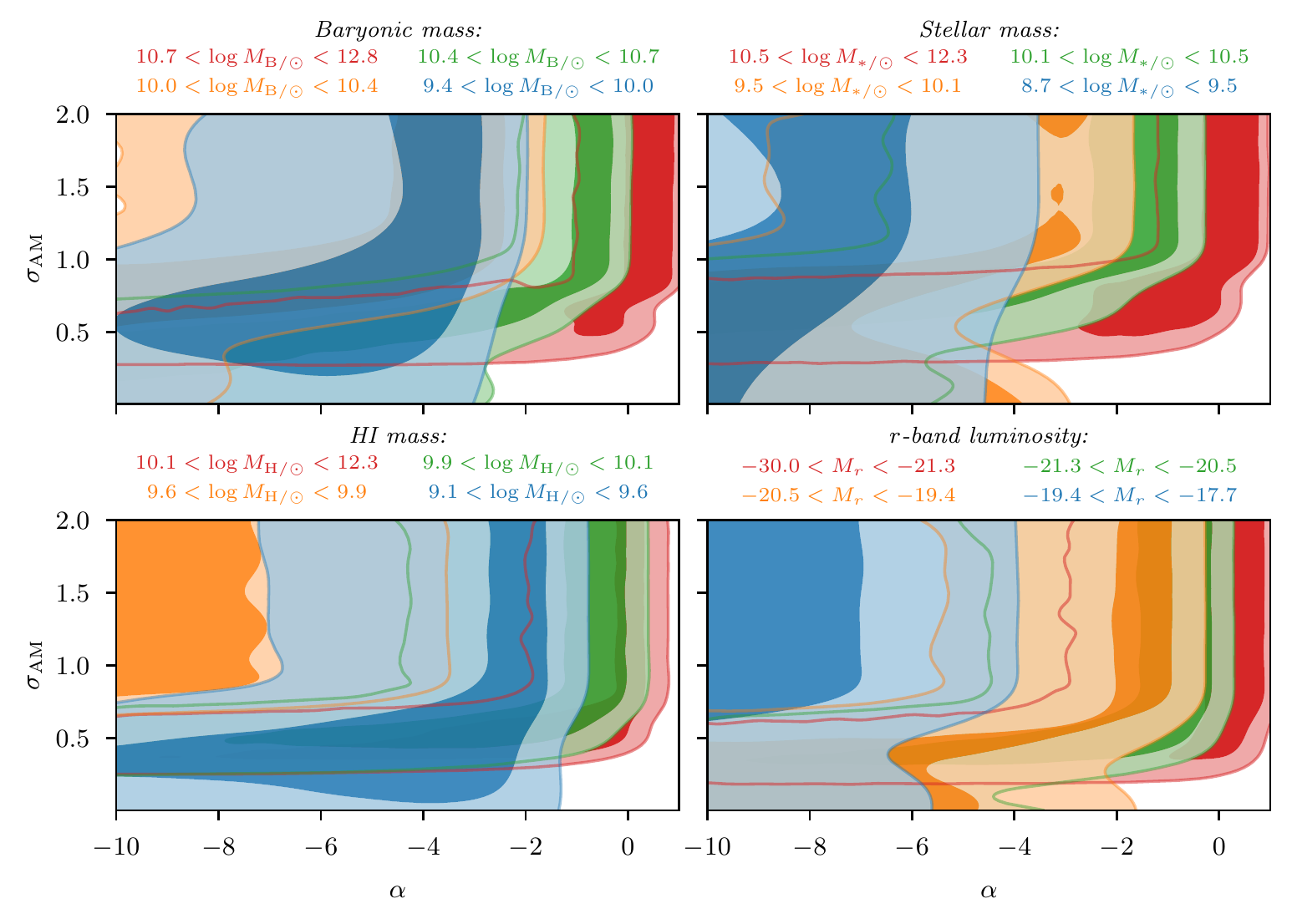}
    \caption{Posteriors on $\alpha$ and $\scatter$ for $\HI$-selected samples from the \matched catalogue. The contours show the minimal areas enclosing $39\%$ and $86\%$ of the probability. The panels are distinguished by the galaxy property used in AM, as indicated.}
    \label{fig:HI_posteriors}
\end{figure*}

We show the posterior contours for baryonic, $\HI$, stellar mass, and $r$-band luminosity-defined samples in Fig.~\ref{fig:HI_posteriors}. As the $\HI$-selected samples are significantly less clustered, the posteriors show a noticeably different dependence on $\alpha$ and $\scatter$. 
In each sample ($M_{\mathrm{B}}$, $M_*$, $M_{\HI}$ or $M_r$) we infer $\alpha < 0$, with the exception of the most massive subsample, with the posterior probability slowly decreasing towards the lower $\alpha$ limit we consider. This corresponds to preferentially occupying haloes with peak mass at present time, i.e. typically excluding subhaloes. At some negative threshold in $\alpha$ the model becomes dominated by haloes that peak at present time, and therefore stops being significantly affected by lowering $\alpha$ further. The exact threshold at which the clustering stops being significantly affected by varying $\alpha$ depends on the subsamples, although there is a systematic trend wherein the high-mass samples have this threshold at higher values of $\alpha$. Therefore, we restrict the uniform prior to $-10 < \alpha < 1$. We do not need to go lower because the likelihood shows little dependence on $\alpha$ for large negative values.

We find substantial support for $\scatter$ of up to $2~\mathrm{dex}$. However, with such $\scatter$ the galaxy--halo connection introduced by the proxy is almost fully randomised, and hence the model contains little physical information. In fact, from the posteriors it is apparent that the ranking of the haloes becomes approximately randomised already at $\scatter$ of $\sim$1 dex, which is when the posteriors become independent of $\scatter$. At this point shuffling the haloes further has little effect on the predicted clustering. We therefore restrict the prior to $\scatter < 2~\mathrm{dex}$. This is typically observed at the higher range of allowed values of $\alpha$, with the exception of the faint $M_r$ and low $M_*$ subsamples, which show strong support for zero scatter at large negative $\alpha$. At such values of $\alpha$ the clustering has already been sufficiently reduced by the proxy, and therefore low $\scatter$, which generally lowers clustering, is sufficient to match the observations

\begin{figure*}
    \centering
    \includegraphics[width=1.0\textwidth]{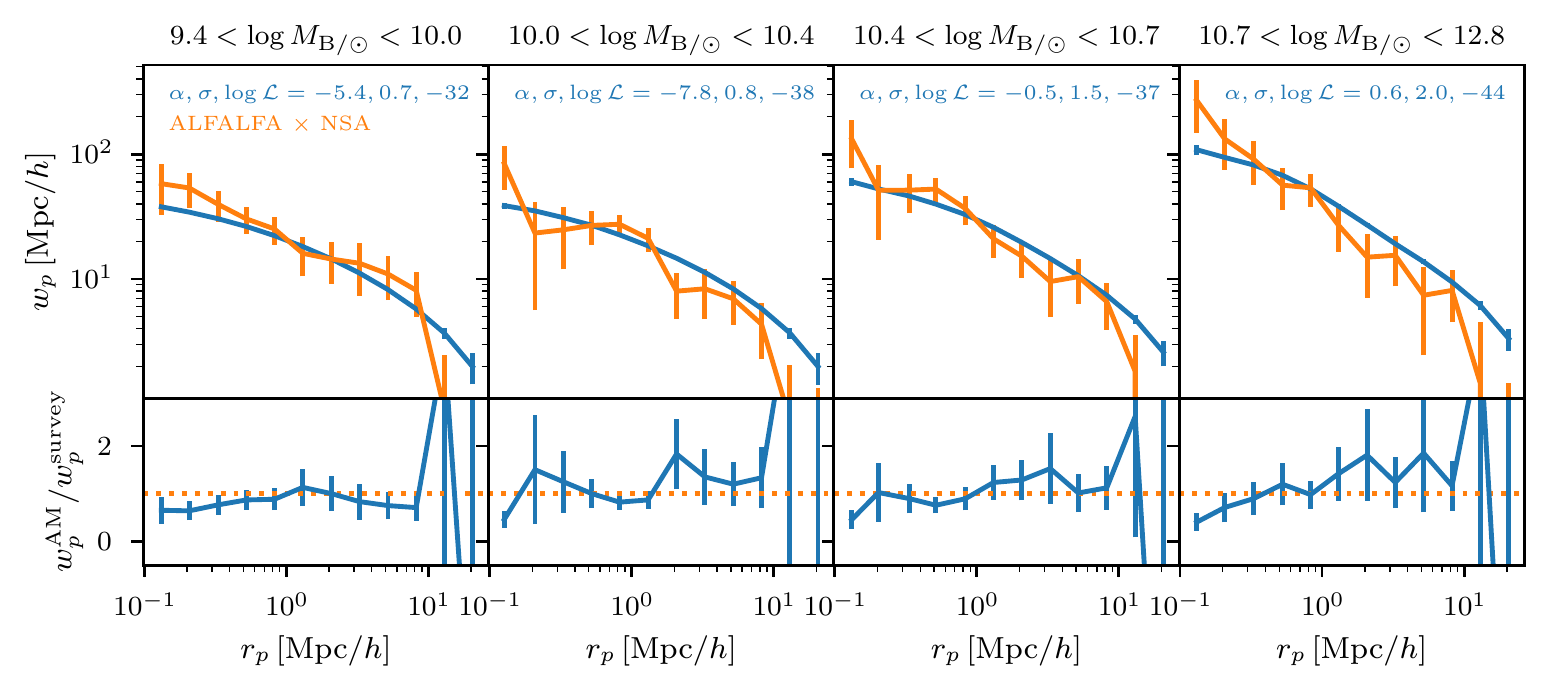}
    \caption{Comparison of the model two-point projected correlation functions for the \matched catalogue from baryonic mass-based \ac{SHAM} vs that observed. The panels show different baryonic mass bins. $\log\mathcal{L}$ denotes the maximum log-likelihood values and the residuals are always taken with respect to the survey's $w_p$.}
    \label{fig:BMF_fits}
\end{figure*}

We also show an example of the reconstructed correlation functions for the $M_\mathrm{B}$ sample in Fig.~\ref{fig:BMF_fits}. These do not show as good agreement as the optical samples, which could be due to two main reasons. First, the proxy is not well-suited to $\HI$-selected galaxies, which is exemplified by the fact that it prefers a very high \ac{SHAM} scatter. Second, the \matched catalogue contains significantly fewer galaxies, and therefore its uncertainty on the two-point projected correlation function overwhelms the uncertainty from \ac{SHAM}. This smears out the variation of the \ac{SHAM} parameters. We note that the goodness of fit in the $M_{\HI}$, $M_*$, or $M_r$ subsamples is marginally worse than with $M_\mathrm{B}$, although a significant conclusion cannot be drawn. We quote the maximum-likelihood points for each sample in Table.~\ref{tab:ML_HIbins}. We do not repeat the Bayes factor analysis for quantifying the tension between subsamples at fixed parameter values because our posteriors are significantly truncated by our prior ranges, which would bias the evidence.

\begin{table*}
\setlength{\tabcolsep}{2.75pt}
\begin{tabular}{llccccccccccccccccccc}
          & & \multicolumn{3}{c}{$M_r:~m_\alpha$}  &  & \multicolumn{3}{c}{$M_*:~m_\alpha$} &  & \multicolumn{3}{c}{$M_{\mathrm{B}}:~m_\alpha$}  &  & \multicolumn{3}{c}{$M_{\mathrm{HI}}:~m_\alpha$} &  & \multicolumn{3}{c}{$M_{\mathrm{B}}:~\zcut$} \\ \cline{3-5} \cline{7-9} \cline{11-13} \cline{15-17} \cline{19-21}
          Percentile range & & $\alpha$   & $\scatter/\mathrm{dex}$ & $\log\mathcal{L}$    &  & $\alpha$  & $\scatter/\mathrm{dex}$ & $\log\mathcal{L}$    &  & $\alpha$   & $\scatter$ & $\log\mathcal{L}$    &  & $\alpha$   & $\scatter/\mathrm{dex}$ & $\log\mathcal{L}$ &  & $\zcut$   & $\scatter/\mathrm{dex}$ & $\log\mathcal{L}$    \\ \cline{1-1} \cline{3-5} \cline{7-9} \cline{11-13} \cline{15-17} \cline{19-21}
$87.5\% \rightarrow 100\%$ & & $0.50$  & $1.5$   & $-49$ &  & $0.33$ & $1.9$   & $-47$ &  & $0.59$  & $2.0$   & $-47$ &  & $-0.89$ & $0.56$  & $-46$  &  & $1.6$ & $0.81$  & $-46$\\
$67.5\% \rightarrow 87.5\%$  & & $-0.45$ & $1.6$   & $-43$ &  & $-1.3$ & $1.8$   & $-43$ &  & $-0.50$ & $1.5$   & $-40$ &  & $-1.0$  & $0.69$  & $-45$ &  & $0.13$ & $0.22$  & $-37$\\
$40.0\% \rightarrow 67.5\%$  & & $-9.8$  & $0.0$   & $-42$ &  & $-9.1$ & $0.0$   & $-40$ &  & $-7.8$  & $0.76$  & $-41$ &  & $-10$   & $0.90$  & $-37$ &  & $0.10$ & $0.19$  & $-38$\\
$10.0\% \rightarrow 40.0\%$   & & $-10$   & $1.9$   & $-37$ &  & $-9.4$ & $0.67$  & $-37$ &  & $-5.4$  & $0.73$  & $-35$ &  & $-3.9$  & $0.45$  & $-37$ &  & $0.32$ & $0.72$  & $-33$ \\ \cline{1-1} \cline{3-5} \cline{7-9} \cline{11-13} \cline{15-17} \cline{19-21}
\end{tabular}
\caption{Maximum-likelihood ($\log\mathcal{L}$) points with corresponding $\alpha$ (or $\zcut$) and $\scatter$ values for the \matched catalogue $r$-band luminosity, stellar mass, baryonic mass and $\HI$ mass-based \ac{SHAM} models. $m_\alpha$ and $\zcut$ in the headings distinguish our two \ac{SHAM} models. The fits using $\zcut$ have marginally higher maximum-likelihood values than using $m_\alpha$.}
\label{tab:ML_HIbins}
\end{table*}

\begin{figure}
    \centering
    \includegraphics[width=1.0\columnwidth]{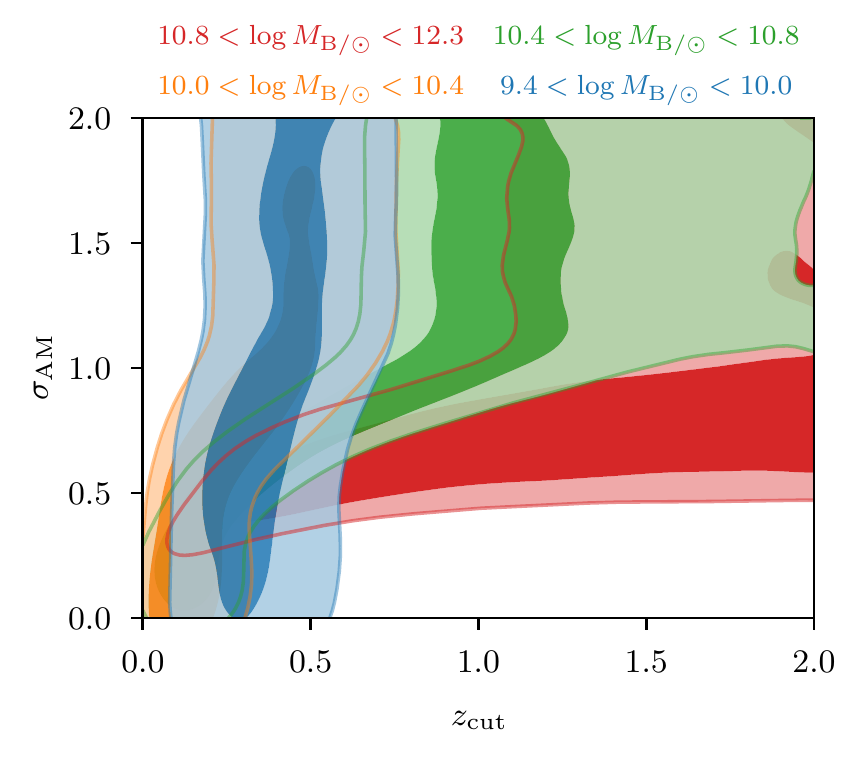}
    \caption{Posteriors on $\zcut$ and $\scatter$ for $\HI$-selected samples for baryonic mass-defined subsamples from the \matched catalogue.
    The posteriors overlap at $\zcut = 0.22^{+0.4}_{-0.2}$ and $\scatter = 0.42^{+0.8}_{-0.2}~\mathrm{dex}$ ($90\%$ confidence intervals).
    The contours show the minimal area enclosing $39\%$ and $86\%$ of the probability.}
    \label{fig:Zcut_posteriors}
\end{figure}

Since in the previous cases the $\HI$-clustering could only be reconstructed at the cost of either very high $\scatter$ or extreme values of $\alpha$, we now introduce a parameter $\zcut$ with the aim of reducing $\scatter$. Ranking haloes by their present virial mass (i.e. fixing $\alpha = 0$), we pre-select only haloes with peak mass redshift lower than $\zcut$. For simplicity we consider only the baryonic mass sample. The resulting posteriors are shown in Fig.~\ref{fig:Zcut_posteriors}. The most massive subsample shows little dependence on $\zcut$, although the $2$nd most massive clearly demonstrates that decreasing $\zcut$ reduces $\scatter$ and the $3$rd and $4$th subsamples only allow $\zcut < 0.5$. The four subsamples' posteriors overlap, with a peak at $\zcut = 0.22^{+0.4}_{-0.2}$ and $\scatter = 0.42^{+0.8}_{-0.2}~\mathrm{dex}$ ($90\%$ confidence intervals). This demonstrates that while ranking haloes by their present mass, $\scatter$ can be reduced by pre-selection on halo formation time (without $\zcut$ we find $\scatter = 0.83^{+0.6}_{-0.5}~\mathrm{dex}$). Regarding the goodness-of-fit of the correlation function we only observe a marginal improvement when adding $\zcut$ (see Table~\ref{tab:ML_HIbins} for the maximum-likelihood points) because the uncertainty of the observed correlation function is significantly larger than that of the \ac{SHAM} mocks.

\section{Discussion}\label{sec:discussion}

\subsection{Interpretation of the results}

Our halo proxy is the product of the present virial mass times and ratio of peak-to-present virial mass to the power of $\alpha$, thereby providing a simple model for halo assembly bias. The proxy reconstructs clustering of all the optically-selected samples we consider to high precision, and also works relatively well for $\HI$-selected galaxies at the price of high scatter in the galaxy--halo connection.

By splitting the samples into lower- and upper-limited subsamples we were able to study the galaxy--halo connection as a function of luminosity and mass. We show that current data is sufficient to place meaningful constraints on this dependence for optically-selected galaxy samples, finding that fainter samples prefer higher $\alpha$ and, especially, $\scatter$ in the \ac{SHAM} prescription. The hypothesis that the \ac{SHAM} parameters are universal is strongly ruled out when comparing the faintest ($M_r > -21.5$ or $M_* < 10^{10.6} M_\odot$) and the brightest subsamples ($M_r < -22.8$ or $M_* > 10^{11.3} M_\odot$). In brighter subsamples ($M_r < -21.5$ or $M_* > 10^{10.6}M_\odot$) our model is consistent with $\alpha \approx 1.2$, corresponding to ranking haloes almost purely by the peak halo mass, and low $\scatter$ of $\sim 0.25~\mathrm{dex}$ suggesting that secondary halo properties play only a minor role in the galaxy--halo connection. On the other hand, the galaxy--halo connection in the fainter subsamples is best fitted with higher $\alpha$ and with $\scatter$ of up to twice that for the brighter or higher-mass subsamples. This indicates that additional variables, not included in our model, are likely relevant in this regime.

We compare \ac{SHAM} based on $r$-band luminosity to stellar mass and show that $r$-band luminosity is strongly favoured for \ac{SHAM} under \ac{NYU} Petrosian photometry and \ac{NSA} elliptical Petrosian photometry, while for \ac{NSA} S\'ersic the results are inconclusive. We also find that the \ac{SHAM} scatter in the $M_*$ model is typically larger. This is perhaps surprising given that the Tully--Fisher relation (TFR) is known to be tighter and more linear when expressed in terms of baryonic mass than luminosity~\citep{McGaugh2000}, which would suggest that galaxy mass is a better indicator of halo properties than magnitude. However, the improvement in the baryonic Tully-Fisher relation (BTFR) is likely due primarily to the inclusion of gas in low-mass, gas-dominated spiral galaxies, and indeed there is little indication that the stellar mass TFR is tighter or more regular than the TFRs defined using luminosities in various bands (e.g.~\citealt{McGaugh_2005, McGaugh_2015}). It is likely that the increased $\scatter$ when using $M_*$ indicates that the stellar masses themselves are imperfect, introducing scatter around the ``true'' $M_*$ values.

We also investigate an $\HI$-selected sample derived by cross-correlating the ALFALFA and \ac{SDSS} data sets. We show that $\alpha > 0$ is strongly excluded in all but the most massive subsamples, which are only consistent with the observations for large $\scatter$ $\gtrsim 0.5~\mathrm{dex}$. We find the galaxy mass or brightness ($M_*,~M_\mathrm{B},~M_{\HI}, M_r$) of $\HI$-selected galaxies not to correlate with the peak halo mass, and only weakly with the present-day virial mass in the most massive galaxy subsamples.

Decreasing $\alpha$ below $0$, which corresponds to assigning bright galaxies preferentially to halos with masses peaking near the present day, typically reduces the scatter in the galaxy--halo connection. This suggests that to describe the galaxy--halo connection in $\HI$ galaxies precisely it is necessary to incorporate other halo properties such as formation time or spin. These properties are far less important for optically-selected samples, where galaxy mass is strongly correlated almost exclusively with halo mass.

To investigate whether $\HI$-selected galaxies are better associated with only a subset of the halo population we tried pre-selecting late-forming haloes before ranking them by their present virial mass. This has a similar effect to lowering $\alpha$, so we do not consider models containing both parameters. We showed that by disregarding haloes that peaked at earlier times the scatter in the galaxy--halo connection can be significantly reduced, to values comparable to the scatter of the optical galaxy--halo connection. This suggests that $\HI$-rich galaxies preferentially reside in halos that formed at lower redshift,
which agrees with the conclusion of~\citet{Guo}.

In the optically-selected samples we find $\alpha \gtrsim 1$, which corresponds to ranking haloes by roughly their peak virial mass before matching them to galaxies. The best-fit values of $\alpha$ increase towards the fainter subsamples. Because increasing $\alpha$ (for $\alpha > 0$) boosts subhaloes, which in turn increases the fraction of satellites in the fainter subsamples, this reflects the fact that the SDSS galaxy sample contains fainter satellite galaxies as well. In sharp contrast, in the $\HI$ samples we find strong support for $\alpha < 0$. In this regime, stripped haloes are down-ranked (assigned to lower mass galaxies), so that the mock catalogues preferentially contain central haloes. This may reflect the fact that satellites are stripped of a large fraction of their $\HI$ gas, in agreement with the lower $\HI$ CF relative to the optical samples where the presence of satellites boosts the small-scale clustering.

\begin{figure}
    \centering
    \includegraphics[width=1.0\columnwidth]{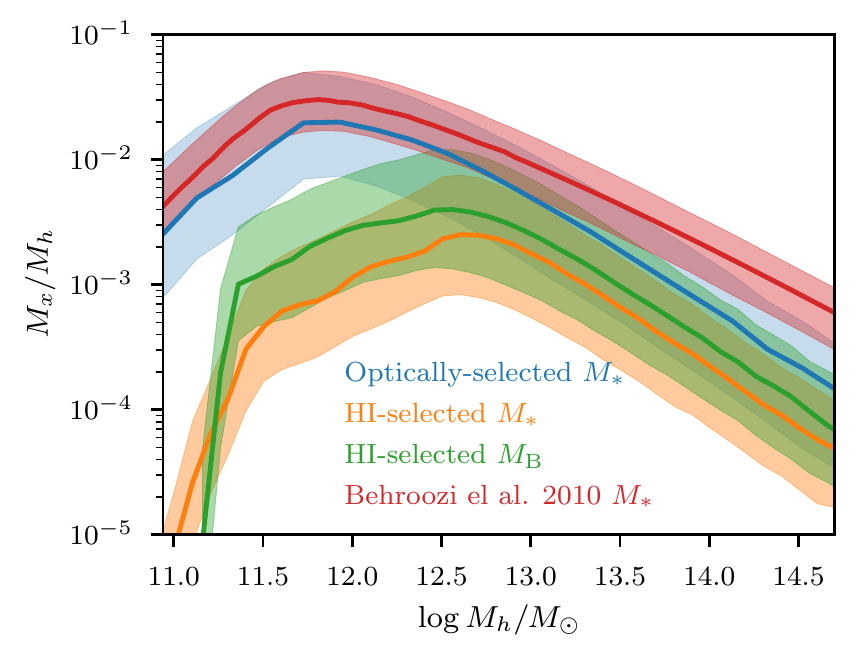}
    \caption{Relation of optically-selected $M_*$, $\HI$-selected $M_*$ and $\HI$-selected $M_{\mathrm{B}}$ to present halo virial mass. The bands denote $1\sigma$ regions, and we compare to the $M_*-M_h$ relation of~\citet{Behroozi2010}.}
    \label{fig:Mh2Mxrelation}
\end{figure}

In Fig.~\ref{fig:Mh2Mxrelation} we show the recovered $M_x-M_h$ relations using the best-fit \ac{SHAM} parameters, where $M_x$ represents optically-selected $M_*$ from \ac{NYU}, $\HI$-selected $M_*$ from the \matched catalogue or $M_\mathrm{B}$ from the \matched catalogue. We observe that the $\HI$-selection results in significantly lower $M_*$ at fixed halo mass, especially below $10^{13} M_\odot$. To obtain these relations we have assumed that the \ac{SHAM} parameters remain constant below the limits of our clustering constraints, despite the fact that we show earlier that these parameters do depend on galaxy brightness. We also extrapolate the mass and luminosity functions towards the faint end with power laws, on which the $M_x-M_h$ relation depends sensitively.

\subsection{Comparison with the literature}

Earlier studies of \ac{SHAM} with \ac{SDSS} galaxies found that values of scatter around $0.2~\mathrm{dex}$ are appropriate for modelling the galaxy--halo connection in bright samples~\citep{Lehmann, Reddick2013}. We find that the galaxy--halo connection is best fitted with scatter of $\sim~0.25$ dex in both $M_r$ and $M_*$ in the bright subsamples ($M_r < -22.8$ or $M_* > 10^{11.3} M_\odot$). In the faintest subsample scatter of $\sim$0.5 dex was found necessary to fit the galaxy--halo connection. The precise boundaries of the faintest subsample considered depend on the specific catalogue, but typically covered the range of $-21.5 < M_r < -20.3$ or $10^{10.0} M_\odot < M_* < 10^{10.6} M_\odot$. Our analysis differs from \citet{Lehmann} in two main respects: 1) we use a different halo proxy, and 2) the subsamples over which we perform \ac{SHAM} have no objects in common.

To investigate which of these differences is more significant we switch to the proxy of~\citet{Lehmann}, which interpolates between the halo virial mass and the maximum circular velocity evaluated at the peak halo mass.
We find little difference in the results as the two proxies are strongly correlated. More significant is the effect of using binned vs thresholded subsamples. Previous works have typically used the latter, with the assumption that galaxies near the lower edge dominate the statistical properties of the sample. However, this assumption is only valid at the bright end where the luminosity or mass function declines sharply. Faint subsamples occupy the regime in which the luminosity function varies only slowly, making their statistical properties dependent on brighter galaxies as well. These subsamples cannot therefore be considered independent when deriving joint constraints on parameters of the galaxy--halo connection. We tested this by removing the upper limits on our subsamples, which resulted in notably different posteriors. In particular the fainter samples' posteriors became skewed towards those of the brighter subsamples, i.e. towards lower scatter.

There have been few past studies of the galaxy--halo connection of $\HI$-selected samples. \citet{Papastergis_HIclust} studied the clustering of $\sim$6,000 $\HI$-selected galaxies from the $30\%$ complete ALFALFA data set, finding no evidence for a dependence of clustering on $\HI$ mass. They argue from this that $\HI$ mass is not strongly correlated with halo mass, and that secondary halo properties such as spin therefore play an important role in the $\HI$ galaxy--halo connection. In contrast, \citet{Guo} found that clustering increases significantly at higher $\HI$ mass in $\sim$16,000 $\HI$-selected galaxies from the $70\%$ complete ALFALFA data set, and that the galaxy--halo connection of these galaxies can be fitted reasonably well by an \ac{SHAM} model that ranks haloes by the peak circular velocity. This required however a pre-selection of late-forming halos according to the parameter $z_{1/2}$, defined as the redshift at which a halo reaches half its peak mass. The $z_{1/2}$ threshold was found to decrease in higher $\HI$ mass subsamples.

We have not pursued here a detailed study of the clustering amplitude dependence on mass. Based on a simple qualitative comparison we note that the clustering of the \matched catalogue shows only marginal dependence on $\HI$ mass, although we find a stronger dependence on the baryonic mass. Similarly to \citet{Papastergis_HIclust} and \citet{Guo} we find that using simply halo mass as the \ac{SHAM} proxy cannot reproduce the observational data due to the poor correlation with the $\HI$-selected galaxy mass, even considering the possibility of scatter in the relation. The proxy we define here, which incorporates information about the halo formation time alongside with the halo mass, can reproduce the clustering reasonably well at the cost of a large scatter. Lastly, and similarly to \citet{Guo}, we also try replacing the proxy by a pre-selection of late-forming haloes in the simulation before performing \ac{SHAM}. This results in a marginally better goodness-of-fit than the peak-to-present halo mass proxy and, more importantly, reduces the \ac{SHAM} scatter. We find that galaxies with $M_{\mathrm{B}} \gtrsim 10^{10.5} M_\odot$ are insensitive to variations of $\zcut$, as most of the highest-mass halos peak in mass at $z=0$, and that in lower-mass galaxies reducing the $\zcut$ threshold decreases the \ac{SHAM} scatter.

Having fitted our \ac{SHAM} model, we also compare the predicted galaxy-to-halo mass relations to that of~\citet{Behroozi2010} in Fig.~\ref{fig:Mh2Mxrelation}. Compared to~\citet{Behroozi2010}, our relation predicts on average lower $M_*$ at a given halo mass, although our model also predicts higher scatter in the relation due to larger $\sigma_\text{AM}$. If we fix $\sigma_\text{AM}=0.2$ dex we obtain only small differences to~\citet{Behroozi2010} due to our differing \ac{SHAM} prescriptions.

\subsection{Systematic uncertainties}\label{sec:systematic_uncertanties}

The two observational inputs to \ac{SHAM} are the halo catalogue from the $N$-body simulation and the galaxy luminosity or mass function. The precision of the simulation depends on the particular $N$-body code, simulation volume and initial conditions. \citet{Schneider} found that the standard $N$-body codes agree to within one percent at $k \leq~1 h\mathrm{Mpc}^{-1}$, the regime of relevance here. The \textsc{rockstar} halo finder was tested in~\citet{Knebe2011}, showing excellent precision in comparison to other halo finders. Moreover, some of the low-mass haloes may not be well resolved or may have been stripped of sufficient mass to fall beneath the simulation resolution limit, potentially biasing the faintest or least massive subsamples we consider~\citep{Contreras2020}. In our halo catalogue the minimum mass halo consists of $\sim 40$ particles, which is insufficient to be well-resolved~\citep{Diemer2015}. Therefore, we chose a conservative lower threshold for the faintest subsamples, such that the majority of matched haloes lie well above the resolution limit. Even in the faintest subsamples, $>95\%$ of haloes contain $>200$ particles, with the median number being around $5000$. To calculate the galaxy luminosity and mass functions for the optically-selected galaxies we used the  $1/V_{\max}$ method, which provides an unbiased, normalised estimate if the survey is complete and contains no significantly under- or over-dense regions~\citep{Efstathiou:1988}. We verified that this is true for our catalogues using the $V/V_{\max}$ test in Sec.~\ref{sec:LF}. With the $\HI$-selected galaxies we used an analogous non-parametric $1/V_{\mathrm{eff}}$ approach (see Sec.~\ref{sec:LF}).

The matching of ALFALFA to \ac{NSA} systematically eliminated low gas mass, optically faint galaxies. Our method for correcting this is described in Sec.~\ref{sec:LF}. Nevertheless, the fact that the \ac{HIMF} in Fig.~\ref{fig:LFs_MFs} begins to turn over at the faint end suggests that this bias correction was not entirely successful. Thus, the ALFALFA results are less reliable at low than high mass, although fully correcting for the exclusion of optically faint galaxies would not alter our qualitative conclusion that regular \ac{SHAM} performs poorly on HI-selected samples.

The survey two-point projected correlation function, used to constrain the \ac{SHAM} parameters, requires as input a uniform distribution of random points matching the survey radial and angular selection criteria. For the radial distribution we used the ``shuffled method'' (see Sec.~\ref{sec:CF}), which provides a less biased estimate than, for example, modelling the survey redshift distribution using a spline fit~\citep{Ross_et_al}. On the other hand, we ignored the surveys' angular selection. In \ac{NYU} the angular selection can be estimated from the spectroscopic completeness of sectors, which yields the proportion of objects with acquired spectra from each sector. We tested applying the \ac{NYU} angular selection corrections and found that it had a negligible effect on the scales we considered. Another possibility is to weight the galaxy pairs when calculating the correlation function, in order to weight areas with different number densities differently (e.g.~\citealt{Feldman}). We found the inclusion of such weighting to have a little effect on our results. This is because the optically-selected \ac{NYU} and \ac{NSA} catalogues contain no significantly under- or over-dense regions. On the other hand, due to the weak clustering of the $\HI$-selected galaxies a precise modelling of the correlation function of the full ALFALFA sample should take such weighting into consideration. Our use of only a subset of the ALFALFA data that is highly complete (Sec.~\ref{sec:data_obs}) means that a weighting scheme is not crucial.

We did not perform model selection on the $\HI$-selected samples because our prior constraints eliminated parts of the posterior with potentially substantial probability density. The constraints on the prior were necessary because the peak-to-present halo mass proxy that we used is not well-suited for $\HI$-selected galaxies, as can be seen from the large preferred scatter values and the fact that the posterior support does not dwindle significantly for very large or small values of $\alpha$ or scatter. More importantly, the large scatter necessitated extrapolation of the mass function at the faint end, which was not constrained by our data, thus potentially biasing the results. Lastly, a possible source of bias in the \matched catalogue is the matching of the $\HI$ to optical sources. This was done by enforcing an angular tolerance of $5\arcsec$ and a line-of-sight distance tolerance of $10~\mathrm{Mpc}$ (Sec.~\ref{sec:data_obs}), which will likely misclassify a small fraction of galaxies. As a qualitative check that this fraction is small, we find few if any obvious outliers in Fig.~\ref{fig:gas_fraction}.

\subsection{Future work}

Most \acl{SHAM} works, including this one, have focused on the local Universe in which galaxy properties may be precisely measured. This has led to a good understanding of the galaxy--halo connection at $z=0$~\citep{Wechsler_2018}. In contrast, a detailed understanding of the galaxy--halo connection as a function of cosmic time is lacking. Therefore, an interesting question is whether our \ac{SHAM} modelling holds at $z > 0$, and if so how its parameters evolve with redshift. This would allow for generating accurate mocks with \ac{SHAM} for upcoming galaxy surveys which extend considerably beyond $z=0$, e.g. Euclid and the Large Synoptic Survey Telescope (LSST)~\citep{Euclid, LSST}.

Ours is one of few studies to explore the $\HI$ galaxy--halo connection in statistical detail. In this and previous works it has been shown that halo assembly bias plays an important and non-trivial role for $\HI$-selected galaxies, with hints that properties like halo formation time or spin are critical components of the galaxy--halo connection. A first step towards designing an \ac{SHAM} model better suited to $\HI$-selected samples would be to introduce a $2\mathrm{D}$ halo proxy, which includes both a primary halo property (mass or velocity) and a secondary halo property like spin, formation time or concentration. We were unable to make strong claims here due to limitations of our proxy and the relatively low number of $\HI$-selected galaxies in our sample, which leads to large uncertainty on the measured correlation function. Future $\HI$ surveys like those conducted by the Square Kilometre Array (SKA)~\citep{SKA} will significantly improve our understanding of this important but poorly-understood aspect of the galaxy--halo connection. In addition, precise inference of the $\HI$--halo connection at the faint end will require the \ac{HIMF} of the matched catalogue to be accurately modelled down to low mass. One way to avoid the downturn that we observe in Fig.~\ref{fig:LFs_MFs} -- at the price of partially specifying the relation a priori -- would be to force the \ac{HIMF} to match the Schechter function fit to an \ac{HIMF} derived from $\HI$ information only, for example~\citet{Jones2018}.

In terms of how the galaxy--halo connection changes with galaxy mass or brightness, there are two possible immediate extensions. The first would be to investigate the variation of the \ac{SHAM} parameters even further into the faint end, for $M_r > -20.3$ or $M_* < 10^{10} M_\odot$. This would provide more definitive evidence for the brightness variation of those parameters and help pin down the precise form of this variation. The second would be to introduce a halo proxy that is explicitly luminosity- or mass- dependent. This would however require a substantial revision of the \acl{SHAM} formalism to introduce a coupling between the galaxy and halo parameters.

The galaxy--halo connection of faint galaxies (beyond the \ac{SDSS} limiting magnitude at distances required for a statistical sample) remains largely unexplored in the context of \acl{SHAM}. The formation and properties of low-mass galaxies serve as important tests of the $\Lambda$CDM model~\citep{Bullock}. We have shown that the faintest subsamples considered ($-21.5 < M_r < -20.3$) require substantially higher scatter in the galaxy--halo connection as well as an altered shape (higher proxy parameter $\alpha$) compared to brighter samples. This suggests that to model the galaxy--halo connection of faint galaxies ($M_r > -19.6$) via \ac{SHAM} a different set of assumptions may be needed, especially concerning satellite galaxies and the importance of halo properties beside virial mass. Our model predicts lower stellar masses at a given halo mass than those in the literature, and we show that the $M_*-M_h$ relation of $\HI$-selected galaxies from the ALFALFA survey is substantially different at the faint end. Extrapolating further, this may help to resolve the missing satellites \citep{Klypin, Moore} and Too Big To Fail \citep{TBTF} problems.

It is also of interest to investigate the galaxy--halo connection as a function of galaxy type. For example, bluer, more gas-dominated galaxies from \ac{SDSS} would be expected to be more similar to $\HI$-selected galaxies. This could be explored by introducing subsamples defined by galaxy type and performing a similar analysis to the one we present here. Moreover, our finding of a significantly larger \ac{SHAM} scatter for fainter samples suggests that assembly bias plays a more important role at the faint end. It would therefore be interesting to investigate whether the inclusion of other secondary halo properties into the galaxy--halo connection can reduce the scatter in this regime.

A good understanding of the galaxy--halo connection is also of use in constraining cosmological parameters from galaxy clustering. While on large scales a simple bias model between galaxy counts and the matter density is sufficient, on smaller scales the details of how galaxies populate halos becomes important. Current surveys typically neglect such scales (e.g.~\citet{DES_1, DES_2}), thereby discarding much pertinent information. The alternative is to use an empirical technique like \ac{SHAM} to generate mock galaxy catalogues and then marginalise over the added parameters in constraining cosmology (e.g.~\citet{Reddick_cosmo}). Systematic effects in the empirical technique must be strictly under control for this approach to be viable.

Lastly, stronger constraints may be obtained by comparing to additional observational data. For example, application of a group finder would allow modelling of the conditional stellar mass functions of satellite and central galaxies~\citep{Reddick2013}.

\section{Conclusions}\label{sec:conc}

We have studied the galaxy--halo connection in both optically and $\HI$-selected galaxies by means of the subhalo abundance matching (SHAM) technique. For the optically-selected galaxies we used the \acl{NYU} with Petrosian magnitudes and stellar masses as well as the \acl{NSA} galaxy catalogue with both the elliptical Petrosian and S\'ersic photometry. We matched galaxies from the ALFALFA survey to the \acl{NSA} to obtain $\HI$-selected galaxies with baryonic mass estimates. We introduced a new parametrised \ac{SHAM} halo proxy which interpolates between the present and peak virial masses, thereby including a simple model for halo assembly bias. The \ac{SHAM} parameters are constrained using measurements of the two-point projected correlation function in bins of luminosity or mass. We show our model to be well-suited to optically-selected galaxies but less so for $\HI$-selected samples where it requires a large scatter, due to the significantly weaker clustering. This likely indicates that unmodelled halo and/or galaxy properties are important in the $\HI$--halo connection. As one way to investigate this we tried pre-selecting only the latest-forming haloes to host $\HI$-rich galaxies, and showed this to be successful in reducing the \ac{SHAM} scatter. Our specific conclusions are as follows:

\begin{itemize}
    \item The best-fit \ac{SHAM} parameters of optically-selected galaxies are dependent on galaxy brightness and mass. Galaxies fainter than $M_r \simeq - 21.5$ or $M_* \simeq 10^{10.6} M_\odot$ require an \ac{SHAM} scatter up to twice as large as the brighter subsamples, as well as systematically larger values of the halo proxy $\alpha$. The strong tension between the \ac{SHAM} parameters in bright and faint subsamples can be seen in Figs.~\ref{fig:posterior_contours} and~\ref{fig:LF_SMF_dependence}. The hypothesis of a universal halo proxy is valid only for $M_r \lesssim - 21.5$ and $M_* \gtrsim 10^{10.6} M_\odot$, for which we quote the best-fit parameters in Table~\ref{tab:combined_posteriors_results}. In the $\HI$-selected samples we also showed that the best-fit \ac{SHAM} parameters depend on galaxy mass (Fig.~\ref{fig:HI_posteriors} and Fig.~\ref{fig:Zcut_posteriors}), and in particular that lower-mass subsamples prefer lower values of $\alpha$. This indicates that halos with mass peaking near the present day, i.e. central haloes, are preferentially populated. However, because our halo proxy is not ideal for such samples we were not able to study this dependence thoroughly (see Sec.~\ref{sec:systematic_uncertanties}).
    \item Different photometric pipelines give marginally different posteriors for the \ac{SHAM} parameters (Fig.~\ref{fig:comparison_nsa2}). This is most significant in the brightest ($M_r < -22.8$ or $M_* > 10^{11.3} M_\odot$) and faintest ($M_r > -21.5$ or $M_* < 10^{10.6} M_\odot$) subsamples. The two selection criteria -- optical and $\HI$ -- strongly affect the preferred \ac{SHAM} parameters, producing mutually exclusive posteriors. Using our peak-to-present virial mass proxy we find that for optically-selected galaxies the proxy parameter is typically $\alpha \approx 1.2$ (Fig.~\ref{fig:posterior_contours}), which corresponds to ranking haloes nearly by their peak mass. On the other hand, for $\HI$-selected galaxies we infer $\alpha \leq 0$, with lower values preferred for fainter subsamples (Fig.\ref{fig:HI_posteriors}). This indicates that halo assembly bias plays out very differently for $\HI$- than optically-selected samples, and is likely more important. To investigate this, we considered a pre-selection of haloes by the peak mass redshift, retaining only haloes that formed before some threshold $\zcut$ and ranking the remaining haloes by their present virial mass. The constraints when combining all subsamples are $\zcut = 0.22^{+0.4}_{-0.2}$ and $\scatter = 0.42^{+0.8}_{-0.2}~\mathrm{dex}$, the latter being a factor of two lower than when $\zcut$ is not included.
    \item The \ac{NYU} Petrosian photometry pipeline maximises the goodness-of-fit of both the $r$-band luminosity and stellar mass-based \ac{SHAM}. In both cases basing \ac{SHAM} on \ac{NSA} S\'ersic photometry recovers the observed clustering worse than either of the two Petrosian photometries.
    \item In the optically-selected samples we compared the goodness-of-fit quality of $r$-band luminosity to stellar mass-based \ac{SHAM}.
    The Bayes factors for the model comparisons are shown in Table~\ref{tab:Mr2M*_comparison}. All the samples we consider prefer $M_r$-based \ac{SHAM}, with Bayes factors of $13$, $190$, and $2200$ for \ac{NYU} Petrosian photometry, \ac{NSA} elliptical Petrosian and S\'ersic photometries respectively. The preference for $M_r$-based over $M_*$-based \ac{SHAM} likely stems from scatter around the ``true'' stellar masses introduced by the $M_*$ definitions that we investigate. For the $\HI$-selected galaxies we considered baryonic mass, $\HI$ mass, stellar and $r$-band luminosity \ac{SHAM}; however, because our model has non-vanishing posterior probability at the prior boundaries in both $\alpha$ and scatter, we were not able to assess quantitatively which galaxy variable is best suited to use in \ac{SHAM} in that case.
    \item The $\HI$-selected samples are substantially less clustered than their optically-selected counterparts across the full range of scales we investigate, and the mass or brightness of $\HI$-selected galaxies is a much poorer indicator of clustering (Fig.~\ref{fig:HI2opt_clustering}). The $\HI$ selection also reduces the stellar mass function at the bright end (Fig.~\ref{fig:LFs_MFs}). Despite the different mass function shape, the weaker clustering still results in substantially different values of the \ac{SHAM} proxy and scatter, and suggests that other galaxy and/or halo properties may be important.
\end{itemize}

We have shown that extrapolating the bright-end galaxy--halo connection to the faint end can lead to biased conclusions: the best-fit bright-end \ac{SHAM} parameters underpredict the clustering at scales $r_p\lesssim 0.75~\mathrm{Mpc}$ by up to $30\%$ at the faint-end. We also found the parameters of \ac{SHAM} and its goodness of fit to be functions of both the photometric reduction method (\ac{NYU}, \ac{NSA}, S\'{e}rsic vs Petrosian magnitudes), and especially between the use of $r$-band luminosity vs stellar mass as input. Lastly, the different selection criteria of the $\HI$ galaxies from the ALFALFA survey suggest that a significant alteration to \ac{SHAM} is required to model the $\HI$--halo connection accurately. A proxy based on peak-to-present virial mass can only recover the $\HI$ clustering reasonably well at the cost of a very large scatter. Therefore, a suitable \ac{SHAM} proxy for $\HI$-selected samples proxy should consider additional halo properties.

\section*{Acknowledgements}
We thank Martha Haynes and Manolis Papastergis for guidance with the ALFALFA data, and Risa Wechsler and Yao-Yuan Mao for comments on the draft.

RS was supported by the Oxford Astrophysics Summer Research Programme. HD is supported by St John's College, Oxford, and acknowledges financial support from ERC Grant No. 693024 and the Beecroft Trust. MGJ was supported by a Juan de la Cierva formaci\'{o}n fellowship (FJCI-2016-29685) from the Spanish Ministerio de Ciencia, Innovaci\'{o}n y Universidades (MCIU) during much of this work. He also acknowledges support from the grants AYA2015-65973-C3-1-R (MINECO/FEDER, UE) and RTI2018-096228-B-C31 (MCIU). This work has been supported by the State Agency for Research of the Spanish MCIU ``Centro de Excelencia Severo Ochoa'' program under grant SEV-2017-0709.

This research made use of the Dark Sky Simulations, which were produced using an INCITE 2014 allocation on the Oak Ridge Leadership Computing Facility at Oak Ridge National Laboratory; we thank the Dark Sky Collaboration for providing access to these simulations. We would like to acknowledge the work of the entire ALFALFA team in observing, flagging, and source extraction for the ALFALFA catalogue.

\section*{Data Availability}
The code underlying this article is publicly available at \href{https://github.com/Richard-Sti/ClusterSHAM}{github.com/Richard-Sti/ClusterSHAM}. The NSA, \ac{NYU} and ALFALFA catalogues are publicly available. All other data used will be shared on reasonable request to the corresponding authors.

\bibliographystyle{mnras}
\bibliography{ref}

\begin{thebibliography}{}
\makeatletter
\relax
\def\mn@urlcharsother{\let\do\@makeother \do\$\do\&\do\#\do\^\do\_\do\%\do\~}
\def\mn@doi{\begingroup\mn@urlcharsother \@ifnextchar [ {\mn@doi@}
  {\mn@doi@[]}}
\def\mn@doi@[#1]#2{\def\@tempa{#1}\ifx\@tempa\@empty \href
  {http://dx.doi.org/#2} {doi:#2}\else \href {http://dx.doi.org/#2} {#1}\fi
  \endgroup}
\def\mn@eprint#1#2{\mn@eprint@#1:#2::\@nil}
\def\mn@eprint@arXiv#1{\href {http://arxiv.org/abs/#1} {{\tt arXiv:#1}}}
\def\mn@eprint@dblp#1{\href {http://dblp.uni-trier.de/rec/bibtex/#1.xml}
  {dblp:#1}}
\def\mn@eprint@#1:#2:#3:#4\@nil{\def\@tempa {#1}\def\@tempb {#2}\def\@tempc
  {#3}\ifx \@tempc \@empty \let \@tempc \@tempb \let \@tempb \@tempa \fi \ifx
  \@tempb \@empty \def\@tempb {arXiv}\fi \@ifundefined
  {mn@eprint@\@tempb}{\@tempb:\@tempc}{\expandafter \expandafter \csname
  mn@eprint@\@tempb\endcsname \expandafter{\@tempc}}}

\bibitem[\protect\citeauthoryear{Abazajian et~al.}{Abazajian
  et~al.}{2009}]{SDSS_DR7}
Abazajian K.~N.,  et~al., 2009, \mn@doi [Astrophys. J. Suppl.]
  {10.1088/0067-0049/182/2/543}, 182, 543

\bibitem[\protect\citeauthoryear{Albareti et~al.}{Albareti
  et~al.}{2017}]{SDSS_DR13}
Albareti F.~D.,  et~al., 2017, \mn@doi [Astrophys. J. Suppl.]
  {10.3847/1538-4365/aa8992}, 233, 25

\bibitem[\protect\citeauthoryear{Baldry, Glazebrook  \& Driver}{Baldry
  et~al.}{2008}]{Baldry:2008ru}
Baldry I.,  Glazebrook K.,   Driver S.,  2008, \mn@doi [Mon. Not. Roy. Astron.
  Soc.] {10.1111/j.1365-2966.2008.13348.x}, 388, 945

\bibitem[\protect\citeauthoryear{{Behroozi}, {Conroy}  \&
  {Wechsler}}{{Behroozi} et~al.}{2010}]{Behroozi2010}
{Behroozi} P.~S.,  {Conroy} C.,   {Wechsler} R.~H.,  2010, \mn@doi [\apj]
  {10.1088/0004-637X/717/1/379}, \href
  {https://ui.adsabs.harvard.edu/abs/2010ApJ...717..379B} {717, 379}

\bibitem[\protect\citeauthoryear{{Behroozi}, {Wechsler}  \& {Wu}}{{Behroozi}
  et~al.}{2013a}]{Behroozi_Rockstar}
{Behroozi} P.~S.,  {Wechsler} R.~H.,   {Wu} H.-Y.,  2013a, \mn@doi [\apj]
  {10.1088/0004-637X/762/2/109}, \href
  {https://ui.adsabs.harvard.edu/abs/2013ApJ...762..109B} {762, 109}

\bibitem[\protect\citeauthoryear{{Behroozi}, {Wechsler}, {Wu}, {Busha},
  {Klypin}  \& {Primack}}{{Behroozi} et~al.}{2013b}]{Behroozi_Consistent}
{Behroozi} P.~S.,  {Wechsler} R.~H.,  {Wu} H.-Y.,  {Busha} M.~T.,  {Klypin}
  A.~A.,   {Primack} J.~R.,  2013b, \mn@doi [\apj]
  {10.1088/0004-637X/763/1/18}, \href
  {https://ui.adsabs.harvard.edu/abs/2013ApJ...763...18B} {763, 18}

\bibitem[\protect\citeauthoryear{{Behroozi}, {Wechsler}, {Hearin}  \&
  {Conroy}}{{Behroozi} et~al.}{2019}]{Universe_machine}
{Behroozi} P.,  {Wechsler} R.~H.,  {Hearin} A.~P.,   {Conroy} C.,  2019,
  \mn@doi [\mnras] {10.1093/mnras/stz1182}, \href
  {https://ui.adsabs.harvard.edu/abs/2019MNRAS.488.3143B} {488, 3143}

\bibitem[\protect\citeauthoryear{Bernardi, Meert, Sheth, Vikram,
  Huertas-Company, Mei  \& Shankar}{Bernardi et~al.}{2013}]{Bernardi}
Bernardi M.,  Meert A.,  Sheth R.~K.,  Vikram V.,  Huertas-Company M.,  Mei S.,
    Shankar F.,  2013, \mn@doi [Mon. Not. Roy. Astron. Soc.]
  {10.1093/mnras/stt1607}, 436, 697

\bibitem[\protect\citeauthoryear{Bishop}{Bishop}{1995}]{10.5555/525960}
Bishop C.~M.,  1995, Neural Networks for Pattern Recognition.
Oxford University Press, Inc., USA

\bibitem[\protect\citeauthoryear{Blanton \& Roweis}{Blanton \&
  Roweis}{2007}]{Kcorrect}
Blanton M.~R.,  Roweis S.,  2007, \mn@doi [Astron. J.] {10.1086/510127}, 133,
  734

\bibitem[\protect\citeauthoryear{Blanton et~al.}{Blanton et~al.}{2005}]{NYU}
Blanton M.~R.,  et~al., 2005, \mn@doi [Astron. J.] {10.1086/429803}, 129, 2562

\bibitem[\protect\citeauthoryear{{Blanton}, {Kazin}, {Muna}, {Weaver}  \&
  {Price-Whelan}}{{Blanton} et~al.}{2011}]{Blanton_subtraction}
{Blanton} M.~R.,  {Kazin} E.,  {Muna} D.,  {Weaver} B.~A.,   {Price-Whelan} A.,
   2011, \mn@doi [\aj] {10.1088/0004-6256/142/1/31}, \href
  {https://ui.adsabs.harvard.edu/abs/2011AJ....142...31B} {142, 31}

\bibitem[\protect\citeauthoryear{{Boylan-Kolchin}, {Bullock}  \&
  {Kaplinghat}}{{Boylan-Kolchin} et~al.}{2012}]{TBTF}
{Boylan-Kolchin} M.,  {Bullock} J.~S.,   {Kaplinghat} M.,  2012, \mn@doi
  [\mnras] {10.1111/j.1365-2966.2012.20695.x}, \href
  {https://ui.adsabs.harvard.edu/abs/2012MNRAS.422.1203B} {422, 1203}

\bibitem[\protect\citeauthoryear{{Bryan} \& {Norman}}{{Bryan} \&
  {Norman}}{1998}]{Bryan_Norman}
{Bryan} G.~L.,  {Norman} M.~L.,  1998, \mn@doi [\apj] {10.1086/305262}, \href
  {https://ui.adsabs.harvard.edu/abs/1998ApJ...495...80B} {495, 80}

\bibitem[\protect\citeauthoryear{{Bullock} \& {Boylan-Kolchin}}{{Bullock} \&
  {Boylan-Kolchin}}{2017}]{Bullock}
{Bullock} J.~S.,  {Boylan-Kolchin} M.,  2017, \mn@doi [\araa]
  {10.1146/annurev-astro-091916-055313}, \href
  {https://ui.adsabs.harvard.edu/abs/2017ARA&A..55..343B} {55, 343}

\bibitem[\protect\citeauthoryear{{Calette}, {Rodr{\'\i}guez-Puebla},
  {Avila-Reese}  \& {del P Lagos}}{{Calette} et~al.}{2021}]{Calette}
{Calette} A.~R.,  {Rodr{\'\i}guez-Puebla} A.,  {Avila-Reese} V.,   {del P
  Lagos} C.,  2021, \mn@doi [\mnras] {10.1093/mnras/stab1788}, \href
  {https://ui.adsabs.harvard.edu/abs/2021MNRAS.tmp.1561C} {}

\bibitem[\protect\citeauthoryear{{Chauhan}, {Lagos}, {Stevens}, {Obreschkow},
  {Power}  \& {Meyer}}{{Chauhan} et~al.}{2020}]{Chauhan2020}
{Chauhan} G.,  {Lagos} C. d.~P.,  {Stevens} A. R.~H.,  {Obreschkow} D.,
  {Power} C.,   {Meyer} M.,  2020, \mn@doi [\mnras] {10.1093/mnras/staa2251},
  \href {https://ui.adsabs.harvard.edu/abs/2020MNRAS.498...44C} {498, 44}

\bibitem[\protect\citeauthoryear{{Chaves-Montero}, {Angulo}, {Schaye},
  {Schaller}, {Crain}, {Furlong}  \& {Theuns}}{{Chaves-Montero}
  et~al.}{2016}]{Chaves-Montero2016}
{Chaves-Montero} J.,  {Angulo} R.~E.,  {Schaye} J.,  {Schaller} M.,  {Crain}
  R.~A.,  {Furlong} M.,   {Theuns} T.,  2016, \mn@doi [\mnras]
  {10.1093/mnras/stw1225}, \href
  {https://ui.adsabs.harvard.edu/abs/2016MNRAS.460.3100C} {460, 3100}

\bibitem[\protect\citeauthoryear{{Conroy}, {Wechsler}  \& {Kravtsov}}{{Conroy}
  et~al.}{2006}]{Conroy2006}
{Conroy} C.,  {Wechsler} R.~H.,   {Kravtsov} A.~V.,  2006, \mn@doi [\apj]
  {10.1086/503602}, \href
  {https://ui.adsabs.harvard.edu/abs/2006ApJ...647..201C} {647, 201}

\bibitem[\protect\citeauthoryear{{Contreras}, {Angulo}  \&
  {Zennaro}}{{Contreras} et~al.}{2020}]{Contreras2020}
{Contreras} S.,  {Angulo} R.,   {Zennaro} M.,  2020, arXiv e-prints, \href
  {https://ui.adsabs.harvard.edu/abs/2020arXiv201206596C} {p. arXiv:2012.06596}

\bibitem[\protect\citeauthoryear{{DES Collaboration} et~al.,}{{DES
  Collaboration} et~al.}{2021}]{DES_1}
{DES Collaboration} et~al., 2021, arXiv e-prints, \href
  {https://ui.adsabs.harvard.edu/abs/2021arXiv210513549D} {p. arXiv:2105.13549}

\bibitem[\protect\citeauthoryear{{Desmond}}{{Desmond}}{2017}]{Desmond_MDAR}
{Desmond} H.,  2017, \mn@doi [\mnras] {10.1093/mnras/stw2571}, \href
  {https://ui.adsabs.harvard.edu/abs/2017MNRAS.464.4160D} {464, 4160}

\bibitem[\protect\citeauthoryear{Desmond \& Wechsler}{Desmond \&
  Wechsler}{2015}]{Desmond_TFR}
Desmond H.,  Wechsler R.~H.,  2015, \mn@doi [Mon. Not. Roy. Astron. Soc.]
  {10.1093/mnras/stv1978}, 454, 322

\bibitem[\protect\citeauthoryear{Desmond \& Wechsler}{Desmond \&
  Wechsler}{2017}]{Desmond_FJR}
Desmond H.,  Wechsler R.~H.,  2017, \mn@doi [Mon. Not. Roy. Astron. Soc.]
  {10.1093/mnras/stw2804}, 465, 820

\bibitem[\protect\citeauthoryear{{Dewdney}, {Hall}, {Schilizzi}  \&
  {Lazio}}{{Dewdney} et~al.}{2009}]{SKA}
{Dewdney} P.~E.,  {Hall} P.~J.,  {Schilizzi} R.~T.,   {Lazio} T.~J.~L.~W.,
  2009, \mn@doi [IEEE Proceedings] {10.1109/JPROC.2009.2021005}, \href
  {https://ui.adsabs.harvard.edu/abs/2009IEEEP..97.1482D} {97, 1482}

\bibitem[\protect\citeauthoryear{{Diemer} \& {Kravtsov}}{{Diemer} \&
  {Kravtsov}}{2015}]{Diemer2015}
{Diemer} B.,  {Kravtsov} A.~V.,  2015, \mn@doi [\apj]
  {10.1088/0004-637X/799/1/108}, \href
  {https://ui.adsabs.harvard.edu/abs/2015ApJ...799..108D} {799, 108}

\bibitem[\protect\citeauthoryear{{Durbala}, {Finn}, {Crone Odekon}, {Haynes},
  {Koopmann}  \& {O'Donoghue}}{{Durbala} et~al.}{2020}]{ALFALFA-SDSS}
{Durbala} A.,  {Finn} R.~A.,  {Crone Odekon} M.,  {Haynes} M.~P.,  {Koopmann}
  R.~A.,   {O'Donoghue} A.~A.,  2020, \mn@doi [\aj] {10.3847/1538-3881/abc018},
  \href {https://ui.adsabs.harvard.edu/abs/2020AJ....160..271D} {160, 271}

\bibitem[\protect\citeauthoryear{{Efstathiou}, {Ellis}  \&
  {Peterson}}{{Efstathiou} et~al.}{1988}]{Efstathiou:1988}
{Efstathiou} G.,  {Ellis} R.~S.,   {Peterson} B.~A.,  1988, \mn@doi [\mnras]
  {10.1093/mnras/232.2.431}, \href
  {https://ui.adsabs.harvard.edu/abs/1988MNRAS.232..431E} {232, 431}

\bibitem[\protect\citeauthoryear{{Eisenstein} et~al.,}{{Eisenstein}
  et~al.}{2001}]{SDSStargeting}
{Eisenstein} D.~J.,  et~al., 2001, \mn@doi [\aj] {10.1086/323717}, \href
  {https://ui.adsabs.harvard.edu/abs/2001AJ....122.2267E} {122, 2267}

\bibitem[\protect\citeauthoryear{{Esmailzadeh}, {Starkman}  \&
  {Dimopoulos}}{{Esmailzadeh} et~al.}{1991}]{Nucleosynthesis}
{Esmailzadeh} R.,  {Starkman} G.~D.,   {Dimopoulos} S.,  1991, \mn@doi [\apj]
  {10.1086/170452}, \href
  {https://ui.adsabs.harvard.edu/abs/1991ApJ...378..504E} {378, 504}

\bibitem[\protect\citeauthoryear{{Feldman}, {Kaiser}  \& {Peacock}}{{Feldman}
  et~al.}{1994}]{Feldman}
{Feldman} H.~A.,  {Kaiser} N.,   {Peacock} J.~A.,  1994, \mn@doi [\apj]
  {10.1086/174036}, \href
  {https://ui.adsabs.harvard.edu/abs/1994ApJ...426...23F} {426, 23}

\bibitem[\protect\citeauthoryear{Giovanelli et~al.}{Giovanelli
  et~al.}{2005}]{Giovanelli_1}
Giovanelli R.,  et~al., 2005, \mn@doi [Astron. J.] {10.1086/497431}, 130, 2598

\bibitem[\protect\citeauthoryear{Giovanelli et~al.}{Giovanelli
  et~al.}{2007}]{Giovanelli_2}
Giovanelli R.,  et~al., 2007, \mn@doi [Astron. J.] {10.1086/516635}, 133, 2569

\bibitem[\protect\citeauthoryear{Guo, Li, Zheng, Mo, Jing, Zu, Lim  \& Xu}{Guo
  et~al.}{2017}]{Guo}
Guo H.,  Li C.,  Zheng Z.,  Mo H.~J.,  Jing Y.~P.,  Zu Y.,  Lim S.~H.,   Xu H.,
   2017, \mn@doi [Astrophys. J.] {10.3847/1538-4357/aa85e7}, 846, 61

\bibitem[\protect\citeauthoryear{{Guo}, {Jones}, {Haynes}  \& {Fu}}{{Guo}
  et~al.}{2020}]{Guo2020}
{Guo} H.,  {Jones} M.~G.,  {Haynes} M.~P.,   {Fu} J.,  2020, \mn@doi [\apj]
  {10.3847/1538-4357/ab886f}, \href
  {https://ui.adsabs.harvard.edu/abs/2020ApJ...894...92G} {894, 92}

\bibitem[\protect\citeauthoryear{{Hamilton} \& {Tegmark}}{{Hamilton} \&
  {Tegmark}}{2004}]{Mangle}
{Hamilton} A.~J.~S.,  {Tegmark} M.,  2004, \mn@doi [\mnras]
  {10.1111/j.1365-2966.2004.07490.x}, \href
  {https://ui.adsabs.harvard.edu/abs/2004MNRAS.349..115H} {349, 115}

\bibitem[\protect\citeauthoryear{Haynes et~al.,}{Haynes
  et~al.}{2018}]{Haynes_2018}
Haynes M.~P.,  et~al., 2018, \mn@doi [The Astrophysical Journal]
  {10.3847/1538-4357/aac956}, 861, 49

\bibitem[\protect\citeauthoryear{{Hogg}}{{Hogg}}{1999}]{Hogg1999}
{Hogg} D.~W.,  1999, arXiv e-prints, \href
  {https://ui.adsabs.harvard.edu/abs/1999astro.ph..5116H} {pp
  astro--ph/9905116}

\bibitem[\protect\citeauthoryear{{Ivezi{\'c}} et~al.,}{{Ivezi{\'c}}
  et~al.}{2019}]{LSST}
{Ivezi{\'c}} {\v Z}.,  et~al., 2019, \mn@doi [\apj] {10.3847/1538-4357/ab042c},
  \href {http://adsabs.harvard.edu/abs/2019ApJ...873..111I} {873, 111}

\bibitem[\protect\citeauthoryear{{Jones}, {Haynes}, {Giovanelli}  \&
  {Moorman}}{{Jones} et~al.}{2018}]{Jones2018}
{Jones} M.~G.,  {Haynes} M.~P.,  {Giovanelli} R.,   {Moorman} C.,  2018,
  \mn@doi [\mnras] {10.1093/mnras/sty521}, \href
  {https://ui.adsabs.harvard.edu/abs/2018MNRAS.477....2J} {477, 2}

\bibitem[\protect\citeauthoryear{{Klypin}, {Kravtsov}, {Valenzuela}  \&
  {Prada}}{{Klypin} et~al.}{1999}]{Klypin}
{Klypin} A.,  {Kravtsov} A.~V.,  {Valenzuela} O.,   {Prada} F.,  1999, \mn@doi
  [\apj] {10.1086/307643}, \href
  {https://ui.adsabs.harvard.edu/abs/1999ApJ...522...82K} {522, 82}

\bibitem[\protect\citeauthoryear{{Knebe} et~al.,}{{Knebe}
  et~al.}{2011}]{Knebe2011}
{Knebe} A.,  et~al., 2011, \mn@doi [\mnras] {10.1111/j.1365-2966.2011.18858.x},
  \href {https://ui.adsabs.harvard.edu/abs/2011MNRAS.415.2293K} {415, 2293}

\bibitem[\protect\citeauthoryear{{Kravtsov}, {Berlind}, {Wechsler}, {Klypin},
  {Gottl{\"o}ber}, {Allgood}  \& {Primack}}{{Kravtsov}
  et~al.}{2004}]{Kravtsov_HOD}
{Kravtsov} A.~V.,  {Berlind} A.~A.,  {Wechsler} R.~H.,  {Klypin} A.~A.,
  {Gottl{\"o}ber} S.,  {Allgood} B.~o.,   {Primack} J.~R.,  2004, \mn@doi
  [\apj] {10.1086/420959}, \href
  {https://ui.adsabs.harvard.edu/abs/2004ApJ...609...35K} {609, 35}

\bibitem[\protect\citeauthoryear{{Kravtsov}, {Vikhlinin}  \&
  {Meshcheryakov}}{{Kravtsov} et~al.}{2018}]{Kravtsov_SMF}
{Kravtsov} A.~V.,  {Vikhlinin} A.~A.,   {Meshcheryakov} A.~V.,  2018, \mn@doi
  [Astronomy Letters] {10.1134/S1063773717120015}, \href
  {https://ui.adsabs.harvard.edu/abs/2018AstL...44....8K} {44, 8}

\bibitem[\protect\citeauthoryear{{Landy} \& {Szalay}}{{Landy} \&
  {Szalay}}{1993}]{Landy_Szalay}
{Landy} S.~D.,  {Szalay} A.~S.,  1993, \mn@doi [\apj] {10.1086/172900}, \href
  {https://ui.adsabs.harvard.edu/abs/1993ApJ...412...64L} {412, 64}

\bibitem[\protect\citeauthoryear{{Laureijs} et~al.,}{{Laureijs}
  et~al.}{2011}]{Euclid}
{Laureijs} R.,  et~al., 2011, arXiv e-prints, \href
  {https://ui.adsabs.harvard.edu/abs/2011arXiv1110.3193L} {p. arXiv:1110.3193}

\bibitem[\protect\citeauthoryear{Lehmann, Mao, Becker, Skillman  \&
  Wechsler}{Lehmann et~al.}{2017}]{Lehmann}
Lehmann B.~V.,  Mao Y.-Y.,  Becker M.~R.,  Skillman S.~W.,   Wechsler R.~H.,
  2017, \mn@doi [Astrophys. J.] {10.3847/1538-4357/834/1/37}, 834, 37

\bibitem[\protect\citeauthoryear{{Li} \& {White}}{{Li} \&
  {White}}{2009}]{Li2009}
{Li} C.,  {White} S. D.~M.,  2009, \mn@doi [\mnras]
  {10.1111/j.1365-2966.2009.15268.x}, \href
  {https://ui.adsabs.harvard.edu/abs/2009MNRAS.398.2177L} {398, 2177}

\bibitem[\protect\citeauthoryear{{Li}, {Kauffmann}, {Fu}, {Wang}, {Catinella},
  {Fabello}, {Schiminovich}  \& {Zhang}}{{Li} et~al.}{2012}]{Li}
{Li} C.,  {Kauffmann} G.,  {Fu} J.,  {Wang} J.,  {Catinella} B.,  {Fabello} S.,
   {Schiminovich} D.,   {Zhang} W.,  2012, \mn@doi [\mnras]
  {10.1111/j.1365-2966.2012.21337.x}, \href
  {https://ui.adsabs.harvard.edu/abs/2012MNRAS.424.1471L} {424, 1471}

\bibitem[\protect\citeauthoryear{{Lu}, {Yang}, {Liu}, {Guo}, {Xu}, {Katsianis}
  \& {Wang}}{{Lu} et~al.}{2020}]{Lu2020}
{Lu} Y.,  {Yang} X.,  {Liu} C.,  {Guo} H.,  {Xu} H.,  {Katsianis} A.,   {Wang}
  Z.,  2020, arXiv e-prints, \href
  {https://ui.adsabs.harvard.edu/abs/2020arXiv200809804L} {p. arXiv:2008.09804}

\bibitem[\protect\citeauthoryear{{Malmquist}}{{Malmquist}}{1920}]{Malmquist1920}
{Malmquist} G.~K.,  1920, Meddelanden fran Lunds Astronomiska Observatorium
  Serie II, \href {https://ui.adsabs.harvard.edu/abs/1920MeLuS..22....3M} {22,
  3}

\bibitem[\protect\citeauthoryear{{Malmquist}}{{Malmquist}}{1922}]{Malmquist1922}
{Malmquist} K.~G.,  1922, Meddelanden fran Lunds Astronomiska Observatorium
  Serie I, \href {https://ui.adsabs.harvard.edu/abs/1922MeLuF.100....1M} {100,
  1}

\bibitem[\protect\citeauthoryear{{Marshall}, {Rajguru}  \& {Slosar}}{{Marshall}
  et~al.}{2006}]{Evidence_test}
{Marshall} P.,  {Rajguru} N.,   {Slosar} A.,  2006, \mn@doi [\prd]
  {10.1103/PhysRevD.73.067302}, \href
  {https://ui.adsabs.harvard.edu/abs/2006PhRvD..73f7302M} {73, 067302}

\bibitem[\protect\citeauthoryear{Martin et~al.,}{Martin et~al.}{2005}]{GALEX}
Martin D.~C.,  et~al., 2005, \mn@doi [The Astrophysical Journal]
  {10.1086/426387}, 619, L1

\bibitem[\protect\citeauthoryear{{Martin}, {Giovanelli}, {Haynes}  \&
  {Guzzo}}{{Martin} et~al.}{2012}]{Martin}
{Martin} A.~M.,  {Giovanelli} R.,  {Haynes} M.~P.,   {Guzzo} L.,  2012, \mn@doi
  [\apj] {10.1088/0004-637X/750/1/38}, \href
  {https://ui.adsabs.harvard.edu/abs/2012ApJ...750...38M} {750, 38}

\bibitem[\protect\citeauthoryear{{McGaugh}}{{McGaugh}}{2005}]{McGaugh_2005}
{McGaugh} S.~S.,  2005, \mn@doi [\apj] {10.1086/432968}, \href
  {https://ui.adsabs.harvard.edu/abs/2005ApJ...632..859M} {632, 859}

\bibitem[\protect\citeauthoryear{{McGaugh} \& {Schombert}}{{McGaugh} \&
  {Schombert}}{2015}]{McGaugh_2015}
{McGaugh} S.~S.,  {Schombert} J.~M.,  2015, \mn@doi [\apj]
  {10.1088/0004-637X/802/1/18}, \href
  {https://ui.adsabs.harvard.edu/abs/2015ApJ...802...18M} {802, 18}

\bibitem[\protect\citeauthoryear{{McGaugh}, {Schombert}, {Bothun}  \& {de
  Blok}}{{McGaugh} et~al.}{2000}]{McGaugh2000}
{McGaugh} S.~S.,  {Schombert} J.~M.,  {Bothun} G.~D.,   {de Blok} W.~J.~G.,
  2000, \mn@doi [\apjl] {10.1086/312628}, \href
  {https://ui.adsabs.harvard.edu/abs/2000ApJ...533L..99M} {533, L99}

\bibitem[\protect\citeauthoryear{{Moore}, {Ghigna}, {Governato}, {Lake},
  {Quinn}, {Stadel}  \& {Tozzi}}{{Moore} et~al.}{1999}]{Moore}
{Moore} B.,  {Ghigna} S.,  {Governato} F.,  {Lake} G.,  {Quinn} T.,  {Stadel}
  J.,   {Tozzi} P.,  1999, \mn@doi [\apjl] {10.1086/312287}, \href
  {https://ui.adsabs.harvard.edu/abs/1999ApJ...524L..19M} {524, L19}

\bibitem[\protect\citeauthoryear{{Moster}, {Somerville}, {Maulbetsch}, {van den
  Bosch}, {Macci{\`o}}, {Naab}  \& {Oser}}{{Moster} et~al.}{2010}]{Moster2010}
{Moster} B.~P.,  {Somerville} R.~S.,  {Maulbetsch} C.,  {van den Bosch} F.~C.,
  {Macci{\`o}} A.~V.,  {Naab} T.,   {Oser} L.,  2010, \mn@doi [\apj]
  {10.1088/0004-637X/710/2/903}, \href
  {https://ui.adsabs.harvard.edu/abs/2010ApJ...710..903M} {710, 903}

\bibitem[\protect\citeauthoryear{{Moster}, {Naab}  \& {White}}{{Moster}
  et~al.}{2018}]{EMERGE2018}
{Moster} B.~P.,  {Naab} T.,   {White} S. D.~M.,  2018, \mn@doi [\mnras]
  {10.1093/mnras/sty655}, \href
  {https://ui.adsabs.harvard.edu/abs/2018MNRAS.477.1822M} {477, 1822}

\bibitem[\protect\citeauthoryear{{Munshi}, {Brooks}, {Applebaum},
  {Christensen}, {Sligh}  \& {Quinn}}{{Munshi} et~al.}{2021}]{Munshi2021}
{Munshi} F.,  {Brooks} A.,  {Applebaum} E.,  {Christensen} C.,  {Sligh} J.~P.,
   {Quinn} T.,  2021, arXiv e-prints, \href
  {https://ui.adsabs.harvard.edu/abs/2021arXiv210105822M} {p. arXiv:2101.05822}

\bibitem[\protect\citeauthoryear{{Navarro}, {Frenk}  \& {White}}{{Navarro}
  et~al.}{1997}]{NFW}
{Navarro} J.~F.,  {Frenk} C.~S.,   {White} S. D.~M.,  1997, \mn@doi [\apj]
  {10.1086/304888}, \href
  {https://ui.adsabs.harvard.edu/abs/1997ApJ...490..493N} {490, 493}

\bibitem[\protect\citeauthoryear{{Norberg}, {Baugh}, {Gazta{\~n}aga}  \&
  {Croton}}{{Norberg} et~al.}{2009}]{Norberg_et_al}
{Norberg} P.,  {Baugh} C.~M.,  {Gazta{\~n}aga} E.,   {Croton} D.~J.,  2009,
  \mn@doi [\mnras] {10.1111/j.1365-2966.2009.14389.x}, \href
  {https://ui.adsabs.harvard.edu/abs/2009MNRAS.396...19N} {396, 19}

\bibitem[\protect\citeauthoryear{{O'Leary}, {Moster}, {Naab}  \&
  {Somerville}}{{O'Leary} et~al.}{2020}]{EMERGE}
{O'Leary} J.~A.,  {Moster} B.~P.,  {Naab} T.,   {Somerville} R.~S.,  2020,
  arXiv e-prints, \href {https://ui.adsabs.harvard.edu/abs/2020arXiv200102687O}
  {p. arXiv:2001.02687}

\bibitem[\protect\citeauthoryear{{Pandey} et~al.,}{{Pandey}
  et~al.}{2021}]{DES_2}
{Pandey} S.,  et~al., 2021, arXiv e-prints, \href
  {https://ui.adsabs.harvard.edu/abs/2021arXiv210513545P} {p. arXiv:2105.13545}

\bibitem[\protect\citeauthoryear{Papastergis, Giovanelli, Haynes,
  Rodríguez-Puebla  \& Jones}{Papastergis et~al.}{2013}]{Papastergis_HIclust}
Papastergis E.,  Giovanelli R.,  Haynes M.~P.,  Rodríguez-Puebla A.,   Jones
  M.~G.,  2013, \mn@doi [Astrophys. J.] {10.1088/0004-637X/776/1/43}, 776, 43

\bibitem[\protect\citeauthoryear{{Petrosian}}{{Petrosian}}{1976}]{Petrosian}
{Petrosian} V.,  1976, \mn@doi [\apjl] {10.1086/182301}, \href
  {https://ui.adsabs.harvard.edu/abs/1976ApJ...209L...1P} {210, L53}

\bibitem[\protect\citeauthoryear{{Reddick}, {Wechsler}, {Tinker}  \&
  {Behroozi}}{{Reddick} et~al.}{2013}]{Reddick2013}
{Reddick} R.~M.,  {Wechsler} R.~H.,  {Tinker} J.~L.,   {Behroozi} P.~S.,  2013,
  \mn@doi [\apj] {10.1088/0004-637X/771/1/30}, \href
  {https://ui.adsabs.harvard.edu/abs/2013ApJ...771...30R} {771, 30}

\bibitem[\protect\citeauthoryear{Reddick, Tinker, Wechsler  \& Lu}{Reddick
  et~al.}{2014}]{Reddick_cosmo}
Reddick R.,  Tinker J.,  Wechsler R.,   Lu Y.,  2014, \mn@doi [Astrophys. J.]
  {10.1088/0004-637X/783/2/118}, 783, 118

\bibitem[\protect\citeauthoryear{{Rodr{\'\i}guez-Puebla}, {Avila-Reese},
  {Firmani}  \& {Col{\'\i}n}}{{Rodr{\'\i}guez-Puebla}
  et~al.}{2011}]{Puebla2011}
{Rodr{\'\i}guez-Puebla} A.,  {Avila-Reese} V.,  {Firmani} C.,   {Col{\'\i}n}
  P.,  2011, \rmxaa, \href
  {https://ui.adsabs.harvard.edu/abs/2011RMxAA..47..235R} {47, 235}

\bibitem[\protect\citeauthoryear{{Ross} et~al.,}{{Ross}
  et~al.}{2012}]{Ross_et_al}
{Ross} A.~J.,  et~al., 2012, \mn@doi [\mnras]
  {10.1111/j.1365-2966.2012.21235.x}, \href
  {https://ui.adsabs.harvard.edu/abs/2012MNRAS.424..564R} {424, 564}

\bibitem[\protect\citeauthoryear{Saintonge, Giovanelli, Haynes, Hoffman, Kent,
  Martin, Stierwalt  \& Brosch}{Saintonge et~al.}{2008}]{Saintonge}
Saintonge A.,  Giovanelli R.,  Haynes M.~P.,  Hoffman G.~L.,  Kent B.~R.,
  Martin A.~M.,  Stierwalt S.,   Brosch N.,  2008, \mn@doi [Astron. J.]
  {10.1088/0004-6256/135/2/588}, 135, 588

\bibitem[\protect\citeauthoryear{{Schmidt}}{{Schmidt}}{1968}]{Schidt:1968}
{Schmidt} M.,  1968, \mn@doi [\apj] {10.1086/149446}, \href
  {https://ui.adsabs.harvard.edu/abs/1968ApJ...151..393S} {151, 393}

\bibitem[\protect\citeauthoryear{Schneider et~al.,}{Schneider
  et~al.}{2016}]{Schneider}
Schneider A.,  et~al., 2016, \mn@doi [JCAP] {10.1088/1475-7516/2016/04/047},
  04, 047

\bibitem[\protect\citeauthoryear{{Sersic}}{{Sersic}}{1968}]{Sersic}
{Sersic} J.~L.,  1968, {Atlas de Galaxias Australes}

\bibitem[\protect\citeauthoryear{{Skibba} \& {Sheth}}{{Skibba} \&
  {Sheth}}{2009}]{Skibba}
{Skibba} R.~A.,  {Sheth} R.~K.,  2009, \mn@doi [\mnras]
  {10.1111/j.1365-2966.2008.14007.x}, \href
  {https://ui.adsabs.harvard.edu/abs/2009MNRAS.392.1080S} {392, 1080}

\bibitem[\protect\citeauthoryear{{Skillman}, {Warren}, {Turk}, {Wechsler},
  {Holz}  \& {Sutter}}{{Skillman} et~al.}{2014}]{DarkSky}
{Skillman} S.~W.,  {Warren} M.~S.,  {Turk} M.~J.,  {Wechsler} R.~H.,  {Holz}
  D.~E.,   {Sutter} P.~M.,  2014, arXiv e-prints, \href
  {https://ui.adsabs.harvard.edu/abs/2014arXiv1407.2600S} {p. arXiv:1407.2600}

\bibitem[\protect\citeauthoryear{Stoughton et~al.}{Stoughton
  et~al.}{2002}]{SDSS_early}
Stoughton C.,  et~al., 2002, \mn@doi [Astron. J.] {10.1086/324741}, 123, 485

\bibitem[\protect\citeauthoryear{Tinker et~al.}{Tinker
  et~al.}{2017}]{Tinker:2016zpi}
Tinker J.~L.,  et~al., 2017, \mn@doi [Astrophys. J.]
  {10.3847/1538-4357/aa6845}, 839, 121

\bibitem[\protect\citeauthoryear{{To}, {Reddick}, {Rozo}, {Rykoff}  \&
  {Wechsler}}{{To} et~al.}{2020}]{Redmapper}
{To} C.-H.,  {Reddick} R.~M.,  {Rozo} E.,  {Rykoff} E.,   {Wechsler} R.~H.,
  2020, \mn@doi [\apj] {10.3847/1538-4357/ab9636}, \href
  {https://ui.adsabs.harvard.edu/abs/2020ApJ...897...15T} {897, 15}

\bibitem[\protect\citeauthoryear{{Vale} \& {Ostriker}}{{Vale} \&
  {Ostriker}}{2004}]{Vale2004}
{Vale} A.,  {Ostriker} J.~P.,  2004, \mn@doi [\mnras]
  {10.1111/j.1365-2966.2004.08059.x}, \href
  {https://ui.adsabs.harvard.edu/abs/2004MNRAS.353..189V} {353, 189}

\bibitem[\protect\citeauthoryear{{Warren}}{{Warren}}{2013}]{2HOT}
{Warren} M.~S.,  2013, arXiv e-prints, \href
  {https://ui.adsabs.harvard.edu/abs/2013arXiv1310.4502W} {p. arXiv:1310.4502}

\bibitem[\protect\citeauthoryear{Wechsler \& Tinker}{Wechsler \&
  Tinker}{2018}]{Wechsler_2018}
Wechsler R.~H.,  Tinker J.~L.,  2018, \mn@doi [Annual Review of Astronomy and
  Astrophysics] {10.1146/annurev-astro-081817-051756}, 56, 435–487

\bibitem[\protect\citeauthoryear{{York} et~al.,}{{York} et~al.}{2000}]{SDSS}
{York} D.~G.,  et~al., 2000, \mn@doi [\aj] {10.1086/301513}, \href
  {https://ui.adsabs.harvard.edu/abs/2000AJ....120.1579Y} {120, 1579}

\bibitem[\protect\citeauthoryear{Zehavi et~al.}{Zehavi
  et~al.}{2005}]{Zehavi_Fiber}
Zehavi I.,  et~al., 2005, \mn@doi [Astrophys. J.] {10.1086/431891}, 630, 1

\bibitem[\protect\citeauthoryear{{Zheng} et~al.,}{{Zheng}
  et~al.}{2005}]{Zheng_HOD}
{Zheng} Z.,  et~al., 2005, \mn@doi [\apj] {10.1086/466510}, \href
  {https://ui.adsabs.harvard.edu/abs/2005ApJ...633..791Z} {633, 791}

\bibitem[\protect\citeauthoryear{{Zwaan}, {Meyer}, {Staveley-Smith}  \&
  {Webster}}{{Zwaan} et~al.}{2005}]{Zwaan}
{Zwaan} M.~A.,  {Meyer} M.~J.,  {Staveley-Smith} L.,   {Webster} R.~L.,  2005,
  \mn@doi [\mnras] {10.1111/j.1745-3933.2005.00029.x}, \href
  {https://ui.adsabs.harvard.edu/abs/2005MNRAS.359L..30Z} {359, L30}

\makeatother
\end{thebibliography}

\label{lastpage}
\end{document}